\documentclass[12pt,thmsa]{article}
\usepackage{amssymb}

\usepackage{graphicx}
\usepackage{sw20aip}

\input{tcilatex}
\begin{document}

\title{The Theory of Caustics and Wavefront Singularities with Physical
Applications }
\author{Juergen Ehlers$^{1}$ and Ezra T. Newman$^{2}$ \\
$^{1}$Max-Planck-Institut fur Gravitationalphysik,\\
Albert Einstein Institut, Schlaatzweg 1,\\
D-14473, Potsdam, Germany\\
$^{2}$Dept. of Physics and Astronomy,\\
Univ. of Pittsburgh, Pgh. PA. 15260 USA}
\date{June 10, 1999}
\maketitle

\begin{abstract}
This is intended as an introduction to and review of the work of V. I.
Arnold and his collaborators on the theory of Lagrangian and Legendrian
submanifolds and their associated maps. The theory is illustrated by
applications to Hamilton-Jacobi theory and the eikonal equation, with an
emphasis on null surfaces and wavefronts and their associated caustics and
singularities.
\end{abstract}

\section{\protect\bigskip Introduction}

The following paper is intended to be an introduction to the theory of
smooth Lagrangian and Legendrian maps from manifolds into manifolds, with a
wide range of examples from physics; Hamilton-Jacobi theory, the theory of
the eikonal equation, wave-fronts, their singularities, caustics, etc. Our
interest in this subject arose out of efforts to understand the beautiful
ideas of V.I. Arnold \cite{30,31,32,39,A6,A5} and his collaborators
concerning the theory of singularities of maps. This effort, in turn, was
motivated originally by a recent reformulation of General Relativity\cite
{34,35} in terms of families of null hypersurfaces which naturally
necessitated a study of the pertinent singularities associated with null
hypersurfaces. Another reason for studying this theory was our interest in
gravitational lens theory\cite{42,38} and Ya. B. Zeldovich's theory of
structure formation in the early universe\cite{Z}.

As Arnold's treatment is much more general than most physicists need or use
and his approach is often quite abstract, we and many colleagues found it
initially difficult to get the essential overall picture of this remarkable
theory. Eventually, to a large extent, the picture did get clarified and we
thought that an elementary presentation, from a physicist's view, of these
ideas might be of use to others who do not have the patience to struggle
through Arnold's beautiful work\cite{30,31,32,39}.

This work is organized as follows. In Sec.II, we discuss the basic geometric
ideas behind the local theory of caustics and wave front singularities\cite
{32,P.L.,A5,33} based on the construction of Lagrangian and Legendrian
submanifolds in phase space via the\textbf{\ }use of \textit{generating
functions} - along with some simple illustrative examples. In Sec.III we
give a particularly instructive example from the theory of the
Hamilton-Jacobi equation. Sec.IV serves to establish various relations of
the eikonal equation and its solutions to (small and large) wave fronts in
arbitrary space-times. After a brief discussion, in Sec.V, of some technical
difficulties which emerge from the use of generating functions in the
detailed implementation of the theory, we proceed, in Sec.VI , to the core
of the method, showing how the concept of \textit{generating} \textit{\
families} (for the construction of Lagrangian and Legendrian submanifolds)
naturally arises and is used to overcome some of the aforementioned
difficulties. Finally, in Sec.VII, we apply this method to wavefronts in ($%
3+1)-$ dimensional space-time, to ensembles of non-interacting particles in
phase space and to gravitational lensing.

The main message we learned from this powerful theory that we wish to convey
to the reader is: in order to describe wave propagation phenomena in space
or space-time - (processes which generically lead to intersections of rays,
focal points or focal lines, sharp edges of wavefronts, infinite densities
or similar ``catastrophes'') - it is advisable to treat the evolution of the
requisite structures not directly in space (space-time), but lift them to a
suitable bundle over space (space-time), evolve them there, and at the end
project the result down into space (space-time). In this way one can
describe the singularities which occur ``downstairs'' in terms of smooth
regular structures ``upstairs''. What might appear at first sight as a
complication turns out to be, in fact, a simplification. While the general
definitions and constructions in Section II are based exclusively on the
differential topology of a manifold, the physical examples employ, as an
additional structure, a (generalized) Hamiltonian which defines a
hypersurface in the bundle space. This leads to the restricted class of
those Lagrangian/Legendrian submanifolds which are contained in those
hypersurfaces and which are ruled by the phase space trajectories. These
particular submanifolds are distinguished from general ones by a certain
rigidity: pieces of them can be continued uniquely by those trajectories. In
accordance with this, in the applications, besides having to satisfy a
certain rank condition, the generating families have to satisfy a first
order partial differential equation, e.g. the Hamilton-Jacobi or eikonal
equation.

Except for the manner of presentation, we do not claim any originality here
and any errors are ours. Though of course there is an overall unity to the
subject, in this elementary treatment we have tried, especially in Sec. II,
to keep separate ideas apart. We have denoted a new idea or topic by a $%
\blacklozenge $. Examples are denoted by a $\spadesuit $.

\section{Lagrangian and Legendrian Submanifolds of Symplectic and Contact
Bundles}

We begin with an arbitrary, $n$-dimensional manifold $M$ to be considered as
a configuration space, with local coordinates $q^{a}$ .

$\blacklozenge $1. Consider the cotangent bundle $T^{*}M$ (or phase space)
over $M$ with fiber coordinates $p_{b}$ , and with associated symplectic
potential $\kappa $ and two-form $\omega $

\begin{equation}
\kappa =p_{a}dq^{a},\quad \;\;\omega =dq^{a}\wedge dp_{a}=-d\kappa .
\label{0}
\end{equation}
We will refer to such a 2n-dimensional symplectic manifold also as $M_{S}$ .

The two-form $\omega $ plays a somewhat similar role in symplectic geometry
as the metric $g$ in Riemannian or Lorentzian geometry. As both $\omega $
and $g$ are non-degenerate, their inverses exist and, respectively, can be
used to lower or raise indices. Metric orthogonality, $g(X,Y)=0,$
corresponds to skew-orthogonality, $\omega (X,Y)=0.$ We shall occasionally
make use of the latter relation.

$\blacklozenge $2. Besides $T^{*}M,$ we shall use $T^{*}M\times \Bbb{R}$
with coordinates $(q^{a},p_{a},u)$ and (by definition) the ``contact'' one
form

\begin{equation}
\alpha =du-p_{a}dq^{a}.  \label{2}
\end{equation}
We call the ``contact manifold'', $(T^{*}M\times \Bbb{R},\alpha ),$ the
``contactification'' of $T^{*}M$ and sometimes use the shorthand $M_{C}$ for
it. A function $U$ on $M$ defines a section of $M_{C}$ (considered as a
bundle with ($n+1)$-dimensional fibers and base $M$) via $u=U(q^{a}),$ $%
p_{a}=\partial _{a}U.$ On such a section, $\alpha =0.$

[Note that though this construction yields a particular example of a contact
manifold, \textit{locally} all contact manifolds can be given this
structure.]

$\blacklozenge $3. We thus have the ``extension'' of the 2n-dimensional
symplectic bundle to the $(2n+1)$-dimensional contact bundle. Alternatively
one can start with a $(2n+2)$-dimensional symplectic bundle and ``reduce''
it to a $(2n+1)$-dimensional contact bundle. [See Remark $5a$, at the end of
this section.]

$\blacklozenge $4. Let $M_{S}$ be a symplectic manifold. An immersed
submanifold, $L$ of $M_{S}$ is called Lagrangian if it is $n$-dimensional
and if the pull-back of $\omega $ to $L$ vanishes. A submanifold of $M_{S}$
is Lagrangian if and only if its tangent spaces are skew-orthogonal to
themselves and have maximal dimension.

$\spadesuit $\ a. As a simple example we can construct a Lagrangian
submanifold in the following manner; Choose a ``generating'' function $%
F=F(q^{a})$, then consider the $n$ $q^{a}$'s as the parameters used to
parametrically describe an $n$-manifold $L$ in the $2n$-dimensional
symplectic space by

\begin{eqnarray}
p_{a} &=&\partial _{a}F(q),  \label{3} \\
q^{a} &=&q^{a}.  \nonumber
\end{eqnarray}

One sees immediately, that on $L,$

\[
\omega =dq^{a}\wedge dp_{a}=(\partial _{a}\partial _{b}F)dq^{a}\wedge
dq^{b}\equiv 0. 
\]

Alternately one could chose $G=G(p)$ as a generating function and define a
Lagrangian submanifold by

\begin{eqnarray*}
q^{a} &=&-\partial ^{a}G(p),\text{ } \\
p_{a} &=&p_{a},
\end{eqnarray*}
with the notation $\partial ^{a}\equiv \partial /\partial p_{a}.$ In
particular, each fiber is a Lagrangian submanifold$.$

In contrast to the first example, Eq.(\ref{3}), the new $L$ will in general
not be a section of the bundle $M_{S}.$ Its projection to $M$ need not be
everywhere a local diffeomorphism. The resulting singularities will occupy
us extensively below.

Other choices include interchanging some of the $p$'s and $q$'s in the
generating function; for example, let $G=G(p_{1},q^{2},....,q^{n})$ with

\begin{eqnarray}
q^{1} &=&-\partial ^{1}G,  \label{4} \\
p_{i} &=&\partial _{i}G,  \nonumber \\
p_{1} &=&p_{1},  \nonumber \\
q^{i} &=&q^{i},  \nonumber
\end{eqnarray}
$i=2,...,n$. $\spadesuit $\ 

\qquad In general there are $2^{n}$ different local representations of
Lagrangian submanifolds in terms of canonical coordinates. To construct them
we divide the set of integers $(1,...,n)$ into two disjoint sets with $%
\widehat{A}$ integers in the first set and $\widehat{J}$ integers in the
second set (with $\widehat{A}+\widehat{J}=n).$ We then choose $\widehat{A}$
different $q^{\prime }s,$ i.e., $(q^{A})$ and $\widehat{J}$ different $%
p^{\prime }s,$ i.e., $(p_{J}).$ A generating function is then chosen as $%
K=K(q^{A},p_{J})$ and a Lagrangian submanifold is given by

\begin{eqnarray}
q^{J} &=&-\partial ^{J}K,  \label{22} \\
p_{A} &=&\partial _{A}K,  \nonumber \\
p_{J} &=&p_{J},  \nonumber \\
q^{A} &=&q^{A}.  \nonumber
\end{eqnarray}

Note that there is never a canonically conjugate pair in the set ($%
q^{A},p_{J}).$

Though it is clear that a submanifold constructed as in Eq.(\ref{22}) is
Lagrangian the converse statement that any Lagrangian submanifold can
locally be constructed in this manner must be proved. We now give a
derivation of this result.

Though the derivation is not difficult, it does get complicated and the
reader might want to skip over the details and simply accept the result or
return to the proof later.

Proof: Let $L$ be a Lagrangian submanifold of $M_{S}$, with $\xi $ a point
on $L$, and let $(p_{a},q^{a})$ be a canonical coordinate system. Since $L$
is an immersed submanifold, there exists a subset of $n$ elements of the set 
$(p_{a},q^{a}),$ say $v^{a}\equiv (p_{i},q^{i^{\prime }}),$ that provides
local coordinates for $L$ near $\xi $ so that $L$ can be represented by $%
w^{a}=f^{a}(v^{b}),$ with $w^{a}$ being the remaining $n$ elements of $%
(p_{a},q^{a}).$

The derivation will consist of two parts; we first show that such a subset $%
v^{a}$ can always be chosen so that it does not contain a canonical pair $%
p_{j},q^{j}$. Then we show that a generating function can be chosen such
that locally $L$ is given by Eq.(\ref{22}).

Instead of giving a proof for arbitrary dimension $n$ of $M$ we take $n=4$
as a representative (and in physics an important) case. The argument will
show how to proceed in general. If the set $v^{a}$ \textit{does} contain a
canonically conjugate pair, we will refer to it as an ``unwanted'' set; if
it does not contain a conjugate pair it will be referred to as a ``desired''
set. Then, it will obviously suffice to prove: If local coordinates on $L$
are given in the first place in one of the ``unwanted'' forms 
\[
(i)\text{ }v^{a}=(p_{1},q^{1},q^{2},q^{3}),\text{ }(ii)\text{ }
v^{a}=(p_{1},q^{1},p_{2},q^{3}),\text{ }(iii)\text{ }
v^{a}=(p_{1},q^{1},p_{2},q^{2}), 
\]
then one can always transform to a system of the ``desired'' form.

Consider case (i). Then, near $\xi $ on $L$, $dp_{1}\wedge dq^{1}\wedge
dq^{2}\wedge dq^{3}\neq 0$. Since $L$ is Lagrangian, we have on $L$, $%
dp_{a}\wedge dq^{a}=0$ (summation convention used). Therefore, 
\[
dp_{1}\wedge dq^{1}\wedge dq^{2}\wedge dq^{3}+dp_{4}\wedge dq^{4}\wedge
dq^{2}\wedge dq^{3}=0; 
\]
and hence the second term, like the first one, does not vanish at $\xi $.
Our assumption (i) implies that, on $L,$ near $\xi $, $%
p_{4}=f(p_{1},q^{1},q^{2},q^{3})$; therefore from

\begin{equation}
dp_{4}=\frac{\partial f}{\partial p_{1}}dp_{1}+\frac{\partial f}{\partial
q^{1}}dq^{1}+\frac{\partial f}{\partial q^{2}}dq^{2}+\frac{\partial f}{%
\partial q^{3}}dq^{3}
\end{equation}
either $\frac{\partial f}{\partial p_{1}}\neq 0$ and $dp_{1}\wedge
dq^{2}\wedge dq^{3}\wedge dq^{4}\neq 0$, or $\frac{\partial f}{\partial q^{1}%
}\neq 0$ and $dq^{1}\wedge dq^{2}\wedge dq^{3}\wedge dq^{4}\neq 0$. From the
implicit function theorem, $p_{4}=f(p_{1},q^{1},q^{2},q^{3})$ can be
inverted so that in the former case, $(p_{1},q^{2},q^{3},q^{4})$ are a
``desired'' set, while in the later case $(q^{1},q^{2},q^{3},q^{4})$ form
the ``desired'' set.

In case (ii), the Lagrange condition gives 
\begin{equation}
dp_{1}\wedge dq^{1}\wedge dp_{2}\wedge dq^{3}+dp_{4}\wedge dq^{4}\wedge
dp_{2}\wedge dq^{3}=0.
\end{equation}
Reasoning as above one eliminates in the second product $dp_{4}$ in favor of 
$dp_{1}$ or $dq^{1}$, obtaining in each case a ``desired'' set.

In the third case (iii) one gets 
\[
dp_{1}\wedge dq^{1}\wedge dp_{2}\wedge dq^{2}+dp_{3}\wedge dp^{3}\wedge
dp_{2}\wedge dq^{2}+dp_{4}\wedge dq^{4}\wedge dp_{2}\wedge dq^{2}=0, 
\]
so either the second or the third term is nonzero at $\xi $. Applying again
the former reasoning to $dp_{3}$ or $dp_{4}$, respectively, one reduces case
(iii) to one of the other cases.

One can thus always transform an ``unwanted'' set to a ``desired'' set.

Accepting now, for an arbitrary $n,$ the existence of a subset $%
(p_{J},q^{A}) $ \textit{without} canonical pairs which provides local
coordinates on $L$ near $\xi $, we use the condition $\omega |_{L}=0$ in the
form 
\[
0=d\kappa =d(p_{J}dq^{J}+p_{A}dq^{A})=d(-q^{J}dp_{J}+p_{A}dq^{A}). 
\]
Thus, there exists locally near $\xi $ a function $K(p_{J},q^{A})$ such
that, on $L$, 
\[
\kappa |_{L}=-q^{J}dp_{J}+p_{A}dq^{A}=dK, 
\]
which means that $L$ is given locally by the $n$ equations, (\ref{22}),
namely 
\begin{equation}
q^{J}=-\partial ^{J}K,\;\;p_{A}=\partial _{A}K.\text{ \qquad }\blacksquare 
\text{Q.E.D.}  \label{22*}
\end{equation}

\begin{itemize}
\item  We note that a generating function can be defined as a potential for
the pull-back of $\kappa $ to a Lagrangian submanifold, $\kappa |_{L}=dK.$

\item  Given a point $\xi $ on $L$, only some the 2$^{n}$ representations
will be valid in its neighborhood. If, for example, $n=2$ and ($q^{1},q^{2}$
) as well as ($q^{1},p_{2}$) are permissible, we have 
\[
dK=p_{1}dq^{1}+p_{2}dq^{2}\text{ and }dG=p_{1}dq^{1}-q^{2}dp_{2}, 
\]

and the change from ($q^{1},q^{2},K$) to ($q^{1},p_{2},G$) is a Legendre
transformation, 
\[
K(q^{1},q^{2})=G(q^{1},p_{2})+q^{2}p_{2},\text{ }p_{2}=\partial _{2}K. 
\]
Globally, $L$ can be ``given'' in terms of an atlas of overlapping charts,
each with a representation of the form, Eq.(\ref{22*}), ``Legendre-related''
in the overlap regions.

\item  Note also that if we have an invertible transformation, $%
y^{a}=Y^{a}(q^{A},p_{J}),$ the Lagrangian submanifold can be parametrized by
the $y^{a}$. In applications, this type of situation, where the Lagrangian
submanifold is parametrized by coordinates other than the $(q^{A},p_{J}),$
is very common. In particular it plays a major role in the discussion of
Sec. VI and VII, on generating families where the parameters $y^{a}$ have a
physical significance.
\end{itemize}

$\spadesuit b.$ We give a simple example of a Lagrangian submanifold
generated by a double-valued ``function'' that is not smooth. The same
submanifold can be generated by a smooth (single-valued) function. Consider $%
\Bbb{R}$ as a configuration space with the generating function

\[
F=\pm q^{3/2} 
\]
so that

\[
p=\partial F/\partial q=\pm \frac{3}{2}q^{1/2}\Rightarrow q=\frac{4}{9}%
p^{2}. 
\]
Note that the second derivative of $F$ at $q=0$ does not exist. The same
Lagrangian submanifold is given by the generating function

\[
G=-\frac{4}{27}p^{3}, 
\]

\[
q=-\partial G/\partial p=\frac{4}{9}p^{2}\Rightarrow F=pq+G=\frac{8}{27}%
p^{3}. 
\]
and is parametrized by $p$ instead of $q$. The projection to the base is
given by $q=\frac{4}{9}p^{2}$ , with the ``critical'' point at $p=0.$ $
\spadesuit $

$\blacklozenge $5. An important issue is the mapping from an $n$-dimensional
Lagrangian submanifold, $L$, to the corresponding $n$-dimensional base space 
$M$. This projection, $\pi ,$ is given locally (from Eq. (\ref{22})) by

\qquad $\qquad \qquad \pi :(q^{A},p_{J})\longmapsto \{q^{A},$ $%
q^{J}=-\partial ^{J}K(q^{A},p_{J})\}$ $.$

For most cases of interest the mapping $\pi $ is, for almost all points, a
diffeomorphism (one-to-one and smooth in both directions). This is the case
whenever $L$ is transversal to the fibers. $L$ may, however, have points
where the Jacobian matrix (the derivative of $\pi )$ is degenerate, i.e.,
has rank lower than $n.$ These are the critical points of $\pi $ which form
the critical set, $CritL;$ the image of $CritL$ in $M$ is the caustic set, $%
\pi (CritL)=CaustL.$ In terms of the preceding representation of $L$ the
critical points are given by

\begin{equation}
det(\partial ^{J}\partial ^{J^{\prime }}K)=0;
\end{equation}
what matters here is the $p_{J}$-dependence of $K.$ Sard's theorem states
that the caustic set has Lebesgue measure zero; the critical set may
however, have positive measure.

Note that the amount by which the rank of the Jacobian matrix, $\frak{J}$ $%
=\pi _{*},$ drops at critical points is equal to the corresponding decrease
in rank for $(\partial ^{J}\partial ^{J^{\prime }}K).$ This integer is an
invariant of $\pi ,$ equal to the dimension of the kernel of $\frak{J},$
i.e., the subspace of the tangent space of $L$ which is annihilated by the
projection. The kernel is given by the solutions $X_{J}$ of ($\partial
^{J}\partial ^{J^{\prime }}K)\,X_{J}=0.$

\begin{itemize}
\item  Given a point $\xi $ on a Lagrangian submanifold one can choose the
coordinate system, ($q^{a}),$ near $\pi (\xi )$ such that a ``desired''
coordinate system has $\widehat{A}=rank\frak{J,}$ $\widehat{J}=$ $dim$ $ker%
\frak{J.}$ The corresponding representation, Eq.(\ref{22*}), contains the
largest number of $q^{\prime }s$ which is possible at $\xi .$ Then, $%
\partial ^{J}\partial ^{J^{\prime }}K=0$ at $\xi ,$ and the kernel of $\frak{%
J}$ is spanned by $\partial ^{J}.$ Such representations are used to give
canonical forms of generating functions near singularities of the Lagrangian
maps.
\end{itemize}

$\spadesuit $.$c$ A simple but important example of a Lagrangian submanifold
will now be constructed and analyzed. It shows why one introduces bundles
and their projections even though one is interested in what is taking place
in $M:$ in the bundle functions are unique and regular and the projection
allows one to control their singularities.

Consider as base manifold $M$ the Euclidean plane $\Bbb{R}^{2}$ with metric $%
\delta _{ab}$ with the associated symplectic manifold $M_{S}=\Bbb{R}^{4},$
with coordinates $(q^{a},p_{a}).$ Now choose a curve $C$ in $M$,
parametrized by $q^{a}=q_{0}^{a}(s)$ in terms of the arc length $s.$ The
(unit) tangent vector, $t^{a}$ and unit normal $n^{a}$ are defined along $C$
by

\begin{equation}
t^{a}\equiv \stackrel{.}{\,q}_{0}^{a}=(t^{1},t^{2})\text{, \qquad }
n^{a}\equiv -\varepsilon ^{a}{}_{b}t^{b}=(-t^{2},t^{1}),  \label{a}
\end{equation}
with a dot denoting differentiation with respect to $s$. The $t^{a}$ and $%
n^{a}$ are related to each other by the (plane) Serret-Frenet equations\cite
{36},

\begin{equation}
\dot{t}^{a}=kn^{a}\text{ and }\dot{n}^{a}=-kt^{a}  \label{b}
\end{equation}
where $k(s)\equiv \delta _{ab}\dot{t}^{a}n^{b}$ and $k^{-1}(s)$ are
respectively the curvature and the radius of curvature of $C$ at $s.$ The
lines in $M$ normal to $C$ are called rays, and their orthogonal curves,
wavefronts.

In the four dimensional space $M_{S}$ of the $(q^{a},p_{a})$, we consider
the two-dimensional surface, $L$, associated with a finite section of $C$
where $k>0,$ by

\begin{eqnarray}
q^{a} &=&q_{0}^{a}(s)+vn^{a}(s),  \label{c} \\
p_{a} &=&\delta _{ab}n^{b}(s)\equiv n_{a}(s),  \label{d} \\
\quad s_{1} &<&s\,<s_{2},\text{ }\quad 0<v<\infty .  \label{e}
\end{eqnarray}
The $v$ and $s$ globally parametrize $L$; different values of $(v,s)$ give
different points of $L$. By direct calculation, one sees that the rank of
the map from $(s,v)$ to ($q^{a},p_{a})$ everywhere equals $2$; this follows
from

\begin{eqnarray*}
dq^{1}\wedge dq^{2} &=&(1-vk)ds\wedge dv, \\
dq^{1}\wedge dp_{2} &=&-(t_{2})^{2}kds\wedge dv, \\
dq^{2}\wedge dp_{1} &=&(t_{1})^{2}kds\wedge dv.
\end{eqnarray*}

\noindent Thus $L$ is a submanifold of $T^{*}M$ on which $(s,v)$ are global
non-canonical coordinates. One sees that if $(1-vk)\neq 0$ then ($%
q^{1},q^{2})$ are preferred coordinates; elsewhere one can use either ($%
q^{1},p_{2}$) or ($q^{2},p_{1}$). Moreover one finds that on $L,$ $\kappa
=p_{a}dq^{a}=dv$ and hence $\omega =dq^{a}\wedge dp_{a}$ $=0,$ and so $L$ is
Lagrangian.

The projection of $L$ to $M$ is given by

\begin{equation}
q^{a}=q_{0}^{a}(s)+vn^{a}(s).
\end{equation}
The Jacobian of this mapping, $(s,p)\rightarrow q^{a}(s,p)$, obtained using
the Serret-Frenet equations, is

\begin{equation}
\left| \frak{J}\right| =\left| 
\begin{array}{cc}
(1-vk)t^{1}(s), & n^{1} \\ 
(1-vk)t^{2}(s), & n^{2}
\end{array}
\right| =1-vk(s).  \label{g}
\end{equation}
Thus the critical set of the projection is the curve on $L$ given by $\left| 
\frak{J}\right| =0$ or

\begin{equation}
v=k(s)^{-1}.  \label{h}
\end{equation}
The caustic is the curve in the base space $M$ given by

\begin{equation}
q_{c}^{a}=q_{0}^{a}(s)+k(s)^{-1}n^{a}(s)  \label{i}
\end{equation}
with, as mentioned earlier, $v=k(s)^{-1}$ the radius of curvature at $s$ of $%
C$\cite{36}. After a brief calculation, one finds that the tangent vector to
the caustic is

\[
\stackrel{.}{q}_{c}^{a}=-\dot{k}k^{-2}n^{a}(s); 
\]
hence the rays are tangent to the caustic.

\begin{remark}
Perhaps a more intuitive way to characterize the caustic directly in $M$ is
to find the points where ``neighboring rays intersect'': 
\[
q_{0}^{a}(s)+vn^{a}(s)=q_{0}^{a}(s+\Delta s)+(v+\Delta v)n^{a}(s+\Delta s) 
\]
leads in the limit $\Delta s\mapsto 0$ to 
\[
\dot{q}_{0}^{a}+v\dot{n}^{a}=-\frac{\Delta v}{\Delta s}n^{a}=\lambda n^{a}, 
\]
and since $\dot{q}_{0}^{a}=t^{a}$ and $\dot{n}^{a}=-kt^{a}$ are orthogonal
to $n^{a}$ this gives $\lambda =0$, and one recovers the earlier caustic
condition, $v=k(s)^{-1}$. Note that defining the caustic in this manner is
equivalent to the search for zero's of Jacobi vector fields, i.e., to points
conjugate to $C$ on rays.
\end{remark}

To further analyze this example we return to the Lagrangian submanifold $%
L,\{ $Eqs.(\ref{c}), (\ref{d}) and (\ref{e})\} and (\textit{first) }define
the ``lifted rays'' by

\begin{eqnarray}
q^{a} &=&q_{0}^{a}(s)+vn^{a}(s),  \label{j} \\
p_{a} &=&n_{a}(s),  \nonumber
\end{eqnarray}
with $s=constant$ and $v$ $=variable$ and (\textit{second) }the ``lifted
wave fronts'' by $v=constant$ and $s$ $=variable.$ The two vector fields
spanning $L$ , $(\partial /\partial v\equiv \widehat{T}_{r}$, $\partial
/\partial s\equiv \widehat{T}_{w}),$ which are tangent respectively to the
lifted rays and wave fronts are expressed in the coordinates ($q^{a},p_{a})$
of $M_{S}$ by

\begin{eqnarray}
\widehat{T}_{r} &=&(n^{a},0)\text{ ,\qquad }\widehat{T}_{r}\cdot \widehat{T}%
_{r}=1,  \label{k} \\
\widehat{T}_{w} &=&(\dot{q}_{0}^{a}+v\dot{n}^{a},\dot{n}%
_{a})=((1-vk)t^{a},-kt_{a})\text{,}  \nonumber \\
\text{ }\widehat{T}_{w}\cdot \widehat{T}_{w} &=&(k)^{2}+(1-vk)^{2}>0\qquad
\end{eqnarray}
with $\qquad $%
\begin{equation}
\widehat{T}_{w}\cdot \widehat{T}_{r}=0  \label{l*}
\end{equation}
where the Serret-Frenet equations have again been used and the scalar
product on $L$ is given by $\hat{U}\cdot \hat{W}\equiv \delta
_{ab}u^{a}w^{b}+\delta ^{ab}\widetilde{u}_{a}\widetilde{w}_{b}$ with $\hat{U}%
=(u^{a},\widetilde{u}_{a}),$ etc.

Evaluating $(\widehat{T}_{r}$, $\widehat{T}_{w})$ on the critical curve,
i.e., Eqs.(\ref{c}) and (\ref{d}) with $v=k(s)^{-1},$ one obtains

\begin{eqnarray}
\widehat{T}_{r} &=&(n^{a},0),\quad \widehat{T}_{r}\cdot \widehat{T}_{r}=1,
\label{m} \\
\widehat{T}_{w} &=&(0,-kt_{a}),\quad \widehat{T}_{w}\cdot \widehat{T}
_{w}=k^{2}>0  \nonumber
\end{eqnarray}
with the tangent vector to the critical curve, 
\begin{equation}
\widehat{T}_{c}=(\frac{-\dot{k}}{k^{2}}n^{a},-kt_{a}),\quad \widehat{T}_{c}%
\widehat{\cdot T}_{c}=k^{2}+(\frac{\dot{k}}{k^{2}})^{2}>0.  \label{n}
\end{equation}
The projections to $M$ of these vector fields are

\begin{equation}
T_{r}^{a}=n^{a},\qquad T_{w}^{a}=(1-vk)t^{a},\qquad T_{c}^{a}=\frac{-\dot{k}%
}{k^{2}}n^{a}  \label{o}
\end{equation}

From Eqs.(\ref{k}), (\ref{l*}), (\ref{m}) and (\ref{n}) we see that the
lifted rays and lifted wave fronts and the critical curve have no stationary
points, (i.e., no zero tangent vectors) while their projections onto $M,$
the wave fronts, do have stationary points (``spikes'' or technically cusps,
[see remark below]) at the caustic, $(1-vk=0).$ $\widehat{T}_{w}$ spans the
kernel of the projection. Note also that at extremals of the curvature $(%
\dot{k}=0,k\neq 0)$ the caustic curve itself has stationary points - again
``spikes'' or cusps - provided that $\ddot{k}\neq 0$ there. See Figs. 1 and
2.

\begin{remark}
To see that indeed the wavefronts and the caustic curve have cusps at the
stationary points of their tangent vectors $T_{w}^{a}$ and $T_{c}^{a}$, we
note the following: if either curve is written as $q^{2}=f(q^{1}$) their
slopes are given, respectively, by $dq^{2}/dq^{1}=t^{2}/t^{1}$ and $%
dq^{2}/dq^{1}=n^{2}/n^{1}$ and hence are well defined at their stationary
points. However, as the stationary points are \textit{smoothly traversed} as
functions of $s$, one sees (by expanding about the stationary point) that
the vectors $T_{w}^{a}$ and $T_{c}^{a}$ point in opposite directions on
either side of the stationary point if $\ddot{k}\neq 0$, giving rise to the
spike appearance.
\end{remark}

These local considerations can be applied to and globalized\cite{Arnoldp1}
for closed convex curves, $C.$ $[$See Chapter 8 of Arnold, \textit{\
Catastrophe Theory, }ref. 2.]

This construction of the normals to a curve in $\Bbb{R}^{2}$ is easily
extended to higher dimensions. For $M=\Bbb{R}^{3}$, one could construct the
normals to an arbitrary 2-surface in $\Bbb{R}^{3}$. See Sec. VII.

From a slightly different physical model as in this example, the same
caustic curve (with the cusps) can easily be observed; it can be seen as the
image on a two-surface, of a point source of light reflected by a distorting
mirror or passing thru a distorting lens. From a different model in $\Bbb{R}
^{3},$ one could visualize the caustics as the ``focusing'' of light rays
from a point source distorted by a mirror, passing through a smoke-filled
room. These caustics would form a ``two-surface''. We will return to the
wave fronts and their singularities shortly via the contact bundle, where
their structure is more natural.$\spadesuit $

$\blacklozenge $6. Turning now to a $(2n+1)$-dimensional contact bundle with
local coordinates $(q^{a},p_{a},u)$ and contact form $\alpha =du-p_{a}dq^{a}$
, we consider the analogue of a Lagrangian submanifold, namely a Legendrian
submanifold, $E$, defined by the requirement that it be an immersed $n$
-dimensional submanifold in the contact manifold and that the contact form
vanishes when pulled back to $E$.

$\spadesuit $\ $d$. A simple example of the construction of an $E$ is to
consider any function $F=F(q^{a})$. With the $q^{a}$ acting as the $n$
parameters for the parametrized form of $E$, $E$ is given by

\begin{eqnarray}
p_{a} &=&\partial _{a}F(q^{a}),  \label{5} \\
u &=&F(q^{a}),  \nonumber \\
q^{a} &=&q^{a}.  \nonumber
\end{eqnarray}
an $n$-dimensional submanifold in the (2n +1)-dimensional contact space.

An alternate form for the construction of an $E$ is to chose the $p_{a}$ as
parameters and take $G=G(p_{a})$ as the generating function. One then has
for the parametrized form of $E$

\begin{eqnarray}
q^{a} &=&-\partial G(p_{a})/\partial p_{a}\equiv -\partial ^{a}G,  \label{5*}
\\
u &=&G(p_{a})-p_{a}\partial ^{a}G,  \nonumber \\
p_{a} &=&p_{a}.  \nonumber
\end{eqnarray}

\begin{itemize}
\item  Note that the $u=u(p_{a})$ is defined via a Legendre transformation
from the $G(p)$.$\spadesuit $
\end{itemize}

The general forms to represent Legendrian submanifolds in terms of a
generating function $G(q^{A},p_{J})$ are - compare with Eq.(\ref{22}) ;

\begin{eqnarray}
q^{J} &=&-\partial ^{J}G,  \label{6**} \\
p_{A} &=&\partial _{A}G,  \label{7**} \\
u &=&G-p_{J}\partial ^{J}G.  \label{8**}
\end{eqnarray}
For other types of parametrization of the Lagrangian and Legendrian
submanifolds, see Sec.IV.

\begin{itemize}
\item  Note that if $E$ is a Legendrian submanifold of $T^{*}M\times \Bbb{R}$%
, its projection into $T^{*}M$ is a Lagrangian submanifold. See the previous
set of equations and Eqs.(\ref{22}).

\item  Also note that if $T^{*}M\times \Bbb{R}$ is considered as a bundle
over $M\times \Bbb{R}$ then its fibers are Legendrian submanifolds.
\end{itemize}

$\blacklozenge $7. The analogue of the Lagrangian mapping of an $n$
-dimensional Lagrangian manifold to the base space $M$, is a ``Legendrian
map'' of the $n$-dimensional Legendrian manifold (a projection) to the $%
(n+1) $ -dimensional space $M\times \Bbb{R}$, i.e., to the original base $M$
times the one dimensional fiber described by $u$, with coordinates ($q^{a},u$%
); it is given locally by Eqs.(\ref{6**}) and (\ref{8**}).

After the projection one then has, in general, an $n$-dimensional
``surface'' - called a ``wave front'' - embedded in an ($n$+1)-dimensional
space. The singularities of this map (where the rank of the Jacobian matrix
drops below $n$) are the wave front singularities.

$\spadesuit e$. A simple but very elucidating example (related to $%
\spadesuit c$) is again to use $\Bbb{R}^{2}$ but now use $r$ and $\phi $ as
coordinates. We first let our configuration space $M$ be a circle $S^{1}$
with coordinate $\phi $, the 2-dimensional symplectic manifold has
coordinates $\phi $ and $p_{\phi }$ . We then identify the contact bundle
coordinate $u$ with the radial coordinate $r$, i.e., we have on our
3-dimensional contact bundle the coordinates, $(\phi ,p_{\phi },r)$. One way
to form a Legendrian submanifold is to take $F=F(\phi )$ and set

\begin{eqnarray*}
p_{\phi } &=&\partial _{\phi }F, \\
\phi &=&\phi , \\
r &=&F(\phi ).
\end{eqnarray*}

Its projection to the $(\phi ,r)$ space is just $r=F(\phi )$, a curve in $%
\Bbb{R}^{2}$ representing a one-dimensional wave front in $\Bbb{R}^{2}$ For
the other forms of the generating function (or for multivalued $F$'s), the
associated wave front would in general have singularities, i.e., singular
points on the front.

We emphasize in this last example that $\Bbb{R}^{2}$ is \emph{not the
configuration space} but is the extension of the configuration space $S^{1}.$
The configuration space is $S^{1}$, the symplectic manifold is the $S^{1}$
with its cotangent vectors and finally the contact bundle is the $\Bbb{R}
^{2} $ with the cotangent vectors over the $S^{1}$.

This construction is easily generalized to higher dimensions. Consider the
configuration space $M$ to be a closed 2-surface in $R^{3}$ with local
coordinates ($\theta $, $\phi $) and

\[
u=r=\sqrt{(x^{2}+y^{2}+z^{2})} 
\]
so that the contact bundle has coordinates $(\theta ,\phi ,p_{\theta
},p_{\phi },r)$. A generating function of the form $F=F(\theta ,\phi )$
yields the Legendrian submanifold

\begin{eqnarray}
p_{\theta } &=&\partial _{\theta }F,  \label{6} \\
p_{\phi } &=&\partial _{\phi }F,  \nonumber \\
\phi &=&\phi ,  \nonumber \\
\theta &=&\theta ,  \nonumber \\
r &=&F(\theta ,\phi ),  \nonumber
\end{eqnarray}
with projection $r=$ $F(\theta ,\phi )$, a two-dimensional wave front in $%
\Bbb{R}^{3}$. Again different forms of the generating function lead to
singularities of the wave front (in general, curves). \{There are other ways
of thinking of these 2-dimensional wave fronts; in the above example we
could think of a null three-surface intersecting a $t=constant$ slice of
Minkowski space-time yielding the wavefront, or having the null surface
intersecting a time-like tube or intersecting a null cone. The latter case
is what occurs in the version of GR where the basic variables are the
light-cone cuts of null-infinity\cite{34,35}.\}

Or consider $M=\Bbb{R}^{3}$, with $(x,y,z,p_{x},p_{y},p_{z},t)$ as the
contact coordinates. Again with $F=F(x,y,z)$ we have the projection into the
four space given by

\begin{equation}
t=F(x,y,z).  \label{7}
\end{equation}

Arnold \cite{31} calls this particular wave front, an example of a ``big
wave front''. In the context of Lorentzian optics (where, of course, the
dynamics determines $F$), Eq.(\ref{7}) describes a null or characteristic
surface. The singularities of the ``big wave front'' are two-dimensional
``surfaces''.

$\spadesuit f.$ In the context of Legendrian submanifolds and maps we
return, for a moment, to example $\spadesuit c$. By contactifying $T^{*}M$
to $T^{*}M\times \Bbb{R},$ i.e., by adding the coordinate $u\equiv t,$ to
the set ($q^{a},p_{a}),$ we obtain the 5-dimensional contact manifold with
coordinates, ($q^{a},p_{a},t).$ The extended base space, $M\times \Bbb{R}%
\equiv \Bbb{R}^{2}\times \Bbb{R},$ can be interpreted as a $(2+1)$
-dimensional flat ``space-time'' with coordinates ($q^{a},t)$. The
Legendrian submanifold, $E$, is constructed from the Lagrangian submanifold $%
L,$ by simply ``adding'' $t=v$ to $L;$ $i.e.,$ $E$ is given by 
\begin{eqnarray}
q^{a} &=&q_{0}^{a}(s)+vn^{a}(s), \\
p_{a} &=&n_{a}(s),  \nonumber \\
t &=&v.
\end{eqnarray}
On $E$ the contact form, $\alpha =dt-p_{a}dq^{a}$ vanishes since $\kappa
=p_{a}dq^{a}=dv$; $\alpha =dt-dv=0.$

The projection of $E$ to the space-time is the ``null surface'', (optical
wavefront or ``big wavefront'') 
\begin{eqnarray}
q^{a} &=&q_{0}^{a}(s)+vn^{a}(s),  \nonumber \\
t &=&v
\end{eqnarray}
possessing a variety of singularities. [See the discussion in $\blacklozenge 
$ 9.] Note that, from Eq.(\ref{i}), the caustic is a non-geodesic null curve
in space-time.. A similar remark applies in higher dimensions$\spadesuit $

$\blacklozenge $8. It must be emphasized that in the case of $\spadesuit a$
and $\spadesuit e$ we have chosen (for simplicity) the generating function
to depend only on the configuration space coordinates $q^{a}$. If the $F,$
so chosen, is a (single-valued) function of the $q^{a}$ there will be no
caustics or wave front singularities. The caustics and wave front
singularities arise from the alternate forms of the generating function or
from ``multivalued functions'' $F(q)$, as illustrated by examples $%
\spadesuit b$ and $\spadesuit f.$

$\blacklozenge $9. One of the most remarkable insights achieved in the
theory of Lagrangian and Legendrian maps is that (in the cases of
configuration spaces, with dimension $n$ $\leq 5$) there is a simple and
complete classification of the associated stable singularities of the maps.
(Stable means that the singularities retain their qualitative,
differential-topological properties under all small perturbations of the
generating function. An example of an unstable singularity is provided by an 
$L$ which is a fiber, $q^{a}=G^{a}(p_{b})=const$.) This classification,
which really lies at the heart of the work of Arnold and coworkers, is based
on the idea of using the allowed fiber preserving canonical coordinate
freedom on the symplectic or contact spaces to put the generating functions
into different inequivalent canonical forms.

We conclude this section with a discussion and a list of the stable
singularities for low dimensions of the base manifold $M.$ We will include
both the singularities of the Lagrangian and Legendrian maps. In particular
we will discuss in detail all the cases of dimension $n$ = 1 and 2 and just
skim over the case of $n$ = 3. The notation used to describe the different
cases, i.e., A$_{i}$ and D$_{i}$ is that of Arnold and arises from the
observation that the classification of singularities is closely related to
the classification of semisimple Lie groups where that notation is used.
Also for simplicity we have excluded from the summary several closely
related cases that differ by signs.

The following material is complicated (though not difficult) and is not
essential for the further understanding of this work. On first reading one
might want to skip it and go straight to the remarks at the end of this
section.

For typographic reasons - the conflict between superscript indices and
powers - we will use the coordinates $(x,y,z)$ for the base space
coordinates and the contact coordinate instead of the customary ($q^{a},u)$
for the remainder of this section.

(a) 1-dimensional $M$ with local coordinate $x$:

A$_{1}:$ The trivial case of a neighborhood of a non-critical point of a
Lagrangian submanifold has the canonical choice of generating function $G=$ -%
$p^{2}$, and thus \{$x=2p,$ $p=p\}$ represents the Lagrangian submanifold
locally.

A$_{2}:$The only other case in 1-dimension is the fold singularity of the
Lagrangian map. Again let $x$ be the coordinate of $M$ and $p$ the momentum
coordinate. A canonical choice for the generating function is $G=-p^{3}.$
The Lagrangian submanifold is given by \{$x=3p^{2},$ $p=p\}$ and the
projection to $M$ is $x=3p^{2}$ which has a ``fold'' singularity at $p=0,$
with rank $\frak{J}=0$. Near the fold, the Lagrangian submanifold covers the
base twice for $x>0$ and not at all for $x<0$; $p$ is a coordinate near the
singularity, $x$ is not.

Extending this to a (three dimensional) contact manifold where the contact
coordinate is $u=y$ then the Legendrian submanifold is given, via $y=G+xp,$
as \{$y=2p^{3},x=3p^{2},p=p\}$ with the Legendrian projection given by \{$%
y=2p^{3},x=3p^{2}\}$ which is a curve in the $(x,y)$ plane having a cusp at $%
p=0\Leftrightarrow (x=0,y=0)$. Alternately the curve can be given by $%
4x^{3}=27y^{2}.$ (Note: the Legendrian projection is a homeomorphism, but
not a diffeomorphism near that point.) This is the most general stable
(local) form that a singular one-dimensional wave front in $\Bbb{R}^{2}$ can
take. This cusp is what we referred to earlier, in $\blacklozenge 5c,$ as a
``spike'' in the wave front. See Fig. 1.

Before proceeding we introduce a convention:

When specifying a generating function $G$ for $n>1$ and we write down only
terms containing $p_{a}$'s; it is to be understood that besides those
variables which are ``visible'', $G$ is always taken to depend trivially on
as many configuration variables as are needed for a ``good'' coordinate
system on $L$. If, e.g., $n=3$ and we write $G=-p_{x}^{3}$, (as in the case $%
n=1,$ $A_{2})$ we mean $G(p_{x},y,z)=-p_{x}^{3}$, so that the corresponding
representation of $L,$ with coordinates ($p_{x},y,z),$ reads 
\[
x=3p_{x}^{2}\quad ,\quad p_{y}=0\quad ,\quad p_{z}=0 
\]
and the projection is given by 
\[
(p_{x},y,z)\rightarrow (3p_{x}^{2},y,z). 
\]

All the action is in the $(x,p_{x})$ pair while the other coordinates are
dummies. This convention is obviously useful; in particular when proceeding
to dimension $n+1$, it is not necessary to list again all the cases for $%
m\leq n$ augmented by dummies. (Note that the amount by which the rank drops
is not affected by dummies.)

(b) 2-dimensional $M=\Bbb{R}^{2}$ with local coordinates ($x,y)$.

Again we have the cases $A_{1}$ and $A_{2}$, augmented as described above.
In the $A_{2}$ Legendrian case, $p_{x}$ and $y$ can be taken as coordinates
on $E$, which is the 2-surface in $(x,y,p_{x},p_{y},z)$-space generated by $%
G=-p_{x}^{3},$ $z=G+xp_{x}$ so that the Legendrian submanifold (with $u=z)$
is given by 
\[
\{x=3p_{x}^{2},\quad y=y,\quad z=2p_{x}^{3},\quad p_{x}=p_{x},\smallskip
\quad p_{y}=0\} 
\]
whose image under projection to $\Bbb{R}^{3}\Bbb{=}M\times \Bbb{R}=(x,y,z)$
is the ``product'' of the algebraic curve $4x^{3}=27z^{2}$ (considered
above) with the $y$-axis. This is a two-dimensional wave front in $\Bbb{R}
^{3}$ that has a ``cusp ridge'' singularity along the $y$-axis. The fold
singularity of the Lagrangian map \textit{becomes} the cusp ridge of the
Legendrian map; see Fig. 3.

A$_{3}:$ The third canonical type of Lagrangian submanifold in $n=2$
dimensions (which has a new form) is given by $G=-(p_{x})^{4}+y(p_{x})^{2}$
yielding 
\begin{equation}
\{x=4(p_{x})^{3}-2yp_{x},\quad y=y,\quad p_{y}=(p_{x})^{2},\quad
p_{x}=p_{x}\}
\end{equation}
with the Lagrangian map $\{x=4(p_{x})^{3}-2yp_{x},$ $y=y\}.$ The critical
points of the Lagrangian submanifold are given by the curve $y=6(p_{x})^{2}$
which when projected to $M$ yields the caustic curve, $\{x=-8(p_{x})^{3},$ $%
y=6(p_{x})^{2}\}$ which is a cusp in the $(x,y)$ plane; the rank $\frak{J}$
drops by one.

[Note that in the two-dimensional plane the only two types of stable
singularities are the folds of the A$_{2}$ maps and the cusps of the A$_{3}$
maps. They can be made physically manifest by the poor focusing of light
from a simple source (e.g., by a point source) onto a plane by a distorting
mirror or glass of water. Their general appearance is often complicated by
the fact that multiple sources often give rise to several different
Lagrangian submanifolds and their respective caustics overlap. Another very
important physical manifestation of these caustics is their appearance in
the ``source'' plane in the theory of gravitational lensing. We will say
more about this in a later section.]

The Legendrian submanifolds obtained from the A$_{3}$ generating function
have the form 
\begin{equation}
\{x=4(p_{x})^{3}-2yp_{x},\text{ }y=y,\text{ }z=3(p_{x})^{4}-yp_{x}{}^{2},%
\text{ }p_{x}=p_{x},\text{ }p_{y}=(p_{x})^{2}\}
\end{equation}
and the Legendrian map to $\Bbb{R}^{3}:(x,y,z),$ given by

\begin{equation}
\{x=4(p_{x})^{3}-2yp_{x},\quad y=y,\quad z=3(p_{x})^{4}-y(p_{x}{})^{2}\}.
\end{equation}

This is a two-dimensional surface (wavefront) in $\Bbb{R}^{3},$ parametrized
by the $(y,p_{x}),$ known as the swallow tail. See Fig. 4. Its critical
points are given by $y=4(p_{x}{})^{2}$ which map to the wavefront
singularities on the curve $\{x=-5(p_{x}{})^{3},$ $y=4(p_{x}{})^{2},$ $%
z=-(p_{x}{})^{4}\}.$\hspace{1in}

(c) 3-dimensional $M$:$(x,y,z);$ there are the same cases as in (a) and (b),
i.e., A$_{1},$A$_{2},$and A$_{3},$ plus three new cases namely

A$_{4}:$ $G=-(p_{x}{})^{5}+z(p_{x}{})^{3}+y(p_{x}{})^{2}$

\noindent and

D$_{4}^{\pm }:$ $G=\mp (p_{x}{})^{2}p_{y}{}+(p_{y}{})^{3}+z(p_{y}{})^{2}.$

The A$_{4}$ caustics (two-surfaces in 3-space) are swallowtails and the D$%
_{4}$ caustics are the so-called elliptic umbilic and hyperbolic umbilic
singularities. (See Arnold\cite{32} for their definitions.) The Legendrian
singularities associated with the A$_{4}$ and D$_{4}$ maps (e.g., the
singularities of the ``big wave front'', the null surfaces in space-time,
are far more complicated. The five singularities listed above, (i.e., A$%
_{2}, $A$_{3},$A$_{4},$D$_{4}^{\pm }),$ applied to the spatial projections
of the big wave-fronts in $(3+1)-$space-time (i.e., the three dimensional
caustic and its singularities) have been treated and shown to be stable.\cite
{HKP}

\begin{remark}
\textbf{: }In both cases A$_{2}$ and A$_{3}$ the rank of the corresponding
Jacobian $\frak{J}$ drops by one at the critical curve. The direction of the
kernel of $\frak{J}$ is tangent to the critical curve only at the cusp point
in the A$_{3}$ case$,$ while it is transverse to the critical curve in the
other case. Similar invariant criteria can be used to characterize the other
singularities. (To see this in the context of lens theory, see Ref.[9]%
\textbf{.}) An advantage of such criteria is that they can be applied
without having to transform to the normal form of the generating function.
In the examples $\spadesuit c$ and $\spadesuit f$ one easily verifies that a
critical point at which $\dot{k}\neq 0$ corresponds to a fold and one at
which $\dot{k}=0$ corresponds to a cusp.
\end{remark}

\begin{remark}
: We want to emphasize that all the different types of caustics in the low
dimensional cases have been observed in optical experiments.\cite{MB}
\end{remark}

\begin{remark}
: In the treatment of symplectic and contact manifolds and their associated
Lagrangian and Legendrian submanifolds that we have given here, we began
with a base space of dimension $n$, then introduced its cotangent bundle $%
T^{*}M$ of dimension 2$n$ (the phase space) and defined Lagrangian
submanifolds as special $n$-dimensional submanifolds in $T^{*}M.$ We then
introduced an additional dimension $\Bbb{R}$, obtaining locally $%
T^{*}M\times \Bbb{R}$ (with the contact coordinate $u$ on $\Bbb{R}$ and
contact form $\alpha =du-p_{a}dq^{a}$ $)$ and thus obtained the ($2n+1)$
-dimensional contact manifold as an ``extension'' (the contactification) of
the phase space with its $n$ dimensional Legendrian submanifolds. We then
considered the projections of the Lagrangian submanifolds onto the $n$%
-dimensional space $M$ (Lagrangian maps) and the projections of the
Legendrian submanifolds onto the $(n+1)$-dimensional space $M\times \Bbb{R}$
(Legendrian maps). We want to point out two aspects of this: \newline
\textbf{a.} We could have started in an alternate way and introduced a
different base space $\widetilde{M}$ (configuration space) of dimension ($%
n+1)$ and its phase space $T^{*}\widetilde{M}$ of dimension ($2n+2$); then
by considering the \textit{projective} cotangent space $PT^{*}\widetilde{M}$
(i.e., non-zero covectors of $\widetilde{M}$ up to scale) we would have
obtained a $(2n+1)$ -dimensional contact space . The contact structure of $%
PT^{*}\widetilde{M}$ arises as follows: the symplectic potential $\widetilde{%
\kappa }$ of $T^{*}\widetilde{M}$ defines on $PT^{*}\widetilde{M}$ a
one-form up to a non-zero factor. Thus the corresponding null vector spaces
(i.e., the annihilators) of the one-forms are unique: they form the ``field
of contact hyperplanes'' on $PT^{*}\widetilde{M}.$ In suitable coordinates \{%
$q^{a},q^{0},p_{a},(p_{0}=-1)\},$ the one-form $dq^{0}-p_{a}dq^{a}$
generates the field of the above contact hyperplanes. Although the intrinsic
and global structures of $PT^{*}\widetilde{M}$ and $M_{C}=T^{*}M\times \Bbb{R%
}$ are different, their dimensions $1+2n,$ are the same and they play the
same role for local considerations. We can choose local coordinates $%
(q^{a},q^{0},p_{a})$ with $p_{0}=-1$ on (part of) $PT^{*}\widetilde{M}$ and
identify them with local coordinates $(q^{a},q^{0}=u,p_{a})$ on $M_{C}$:
Then the hypersurface elements given by $dq^{0}-p_{a}dq^{a}=0$ correspond to
those given by $du-p_{a}dq^{a}=0$. Therefore, \textit{if } the objects of
interest are these elements and not the 1-forms themselves, one may \textit{%
locally} work with either $PT^{*}\widetilde{M}$ or $M_{C}$ and their
Legendrian submanifolds and projections. This applies in particular to the 
\textit{local} study of null hypersurfaces in space-time, the latter being
represented either as $\widetilde{M}$ or $M\times \Bbb{R}$. For their 
\textit{global} analysis, $PT^{*}\widetilde{M}$ is the appropriate setting. 
\newline
\textbf{b.} Though it is often natural to think of the configuration space $%
M $ as the physical space (e.g., when one discusses caustics of families of
light rays), nevertheless it is equally often useful to think of the $(n+1)$
space $M\times \Bbb{R}$ as the physical space or space-time (e.g., when
studying wavefronts and their singularities.) Sometimes the relations
between these two interpretations can get quite confusing. Depending on the
physical situation the relationships could be quite different.
\end{remark}

\begin{remark}
: In this section we have mainly tried to give an exposition of the
mathematics of Lagrangian and Legendrian submanifolds and their maps to $M$
and $M\times \Bbb{R},$with occasional digressions to their connections with
physics. In particular we have just explained that a variety of generating
functions can be used to obtain a variety of Lagrangian and Legendrian
submanifolds and their maps \textit{but we have essentially avoided
describing how they are to be physically chosen. }We have done this for two
reasons; pedagogically we thought it best to first describe the mathematics
and second because the variety of different physical uses could in their own
way be confusing. We think that there is however one essential idea that is
common to all (or at least most) uses; when a Lagrangian (Legendrian)
submanifold is chosen it should be thought of as a particular \textit{\
ensemble} of states of a physical system, i.e. for each point of the
submanifold there is a particle (photon or light-ray in the case of
geometric optics) with a particular position and momentum. Sometimes the
submanifold will be thought of as representing the initial conditions for
the ensemble, other times it will represent the evolution of a smaller
ensemble. Later we will discuss these ideas in the context of the
Hamilton-Jacobi equation with an emphasis on the \textit{eikonal equation},
i.e., the massless Hamilton-Jacobi equation, and the beautiful theory of 
\textit{\ generating families}.
\end{remark}

\begin{remark}
:We point out a fascinating historic fact that seems not to be well known;
Einstein\cite{Einstein} as early as 1917 in his investigations (involving
both improvements and serious criticisms) of the Sommerfeld-Epstein quantum
rules, very clearly came across the existence of Lagrangian submanifolds; he
clearly saw that generating functions of the form $G(q^{a})$ could, in
general, only be given locally or as multivalued functions and that in
regions there could be lower dimensional subspaces of critical points.
\end{remark}

\section{An Example from Dynamics}

In this section we will give a simple but very illustrative example of the
physical use of the mathematical ideas described in the previous section.
The example comes from a verbal discussion given by Arnold in reference [2]
and worked out in reference [10].

Consider a one-dimensional configuration space (and its associated phase
space) with a free particle Hamiltonian,

\qquad \qquad \qquad \qquad $H=\frac{1}{2}p^{2}.$

We want to treat the evolution of an ensemble of free particles, or
equivalently a pressureless fluid, with some given initial conditions. This
problem can be treated either directly via the particle motions, ($%
x=x_{0}+pt),$ or via the Hamilton-Jacobi (H-J) equation\qquad \qquad \qquad
\qquad 
\begin{equation}
\frac{\partial S}{\partial t}+\frac{1}{2}(\frac{\partial S}{\partial x}
)^{2}=0.  \label{j1}
\end{equation}

We select the latter method since it illustrates the material of both the
previous and later sections.

We choose, as an example, the following \textit{initial }momentum
distribution \qquad \qquad $\qquad \qquad \qquad $%
\begin{equation}
p=\frac{1}{1+x^{2}},  \label{e0}
\end{equation}
defining a Lagrangian submanifold in the\textbf{\ }$(x,p)$\textbf{\ }phase
space. Since $p=\frac{\partial S}{\partial x},$ we obtain the initial value
of the action function (or velocity potential)\qquad \qquad \qquad \qquad 
\begin{equation}
S_{0}=tan^{-1}x.  \label{j2}
\end{equation}
Simply from the physical situation we expect the faster moving particles
eventually to overtake the slower ones and that the single valued momentum
field, Eq.(\ref{e0}), should change to a multivalued one. At points where
the ``multi-valuedness'' starts or ends, we expect to find the focusing or
caustic points of the projection map. [See fig.5]

Using, from Eq.(\ref{e0}), $x\equiv x_{0}=\pm \sqrt{p^{-1}-1}$ for the
initial position as a function of the momenta, the equations for the
particle motions, namely 
\begin{equation}
g_{\pm }=x-pt-x_{0}\equiv x-pt\mp \sqrt{p^{-1}-1}=0\text{,}  \label{j0}
\end{equation}
implicitly define a function $p(x,t)$ in the strip $0\leq t<\frac{8}{9}\sqrt{%
3}\equiv t_{c},$ (for the meaning of $t_{c},$see below), $-\infty <x<\infty
. $ (To see this, consider $g_{\pm }$ $=0$ for fixed $(x,t)$ in $0<p\leq 1.$
\ Then $g_{+}=0$ ($g_{-}=0)$ has a unique solution for $p$ if $x\geq t$ $%
(x\leq t),$ and for $x=t$ the solutions coincide, $p(x,x)=1.)$ Using this
function, $p(x,t)=\frac{\partial S(x,t)}{\partial x},$ we can write down the
solution of the time dependent H-J equation,(\ref{j1})

\begin{equation}
S(x,t)=\frac{1}{2}p^{2}t+\tan ^{-1}(x-pt)  \label{e1j}
\end{equation}
where $p=p(x,t)$ is that defined above in Eq.(\ref{j0}), in the
aforementioned strip. Note that this solution can be directly obtained by
integrating the H-J equation with the initial data, Eq.(\ref{j2}); see
Secs.IV and VII.

At \textit{all times t,} the particle states $(x,p)$ are given by the
(cubic) algebraic curve$,$ (obtained from $g_{+}g_{-}=0),$

\begin{equation}
p(x-pt)^{2}+p-1=0.  \label{j4}
\end{equation}
which represents a family $L(t)$ of Lagrangian submanifolds in the $(x,p)$
phase space. For $0\leq t<t_{c},$ $L(t)$ is generated by $S(x,t),$

\begin{equation}
p=\frac{\partial S(x,t)}{\partial x},\text{ }x=x
\end{equation}
with a trivial projection (diffeomorphism). But for $t>t_{c},$ there is a
time-dependent open interval $x_{1}<x<x_{2}$ where Eq.(\ref{j4}) has three
solutions, $p_{i},$ i.e., $L(t)$ has two folds; see Fig.4. Near the folds $S$
does not generate $L(t).$ However near the folds we can introduce an
alternative generating function $G(p,t),$ obtained, first, for $t<t_{c}$ by
the Legendre transformation

\begin{equation}
G(p,t)=S(x,t)-px=-\frac{1}{2}p^{2}t+\tan ^{-1}(\sqrt{p^{-1}-1})-p\sqrt{
p^{-1}-1}  \label{j5}
\end{equation}
in the domain $x>t$ and then continued to $t\geq t_{c}.$ Then $L(t),$
including the critical points, is given by

\begin{equation}
x=-\frac{\partial G}{\partial p}=pt+\sqrt{p^{-1}-1},\text{ }p=p  \label{j6}
\end{equation}
Dropping the trivial part, $p=p,$ Eq.(\ref{j6}) gives the projection onto
the $x$-space.

The critical points (where the folds are) are given by those values of $p$
where

\begin{equation}
\frac{dx}{dp}=0=t-\frac{1}{2p^{2}\sqrt{p^{-1}-1}}.  \label{e3}
\end{equation}
Eq.(\ref{e3}) can be rewritten as$\qquad \qquad \qquad \qquad $%
\begin{equation}
f(p,t)\equiv p^{4}-p^{3}+\frac{1}{(2t)^{2}}=0.
\end{equation}
Thought of as a function of $p,$ $f(p,t)$ has a minimum at $p=3/4,$ and a
point of inflection with a double extremum at $p=0.$ One sees that at the 
\textit{minimum}, for $t<t_{c}\equiv \frac{8}{9}\sqrt{3},$ $f(3/4,t)$ is
positive and hence $f(p,t)$ does not vanish for any $p$ while for $t>\frac{8%
}{9}\sqrt{3}$ there are always two solutions; one solution lies between 0
and $3/4$ while the other lies between $3/4$ and $1$. As $t=>$ $\infty ,$
the two roots approach respectively $0$ and $1.$ The values of $x,$ in Eq.(%
\ref{j6} ), associated with these critical points are the caustics which
move to the right along the x-axis with increasing $t$.

As we mentioned earlier, physically we can think of this example as
representing an ensemble of free particles moving to the right with an
initial momentum distribution. After some time the faster ones catch up with
the slower ones and the distribution becomes triple valued with two
caustics. If there had been some initial smooth density $\rho =\rho
_{0}(x_{0}),$ the density at later times is given by $\rho (x,t)=\rho
_{0}(x_{0})(dx_{0}/dx),$ where the initial position $x_{0}$ as a function of 
$x$ and $t$ is obtained by inserting the $p$ from Eq.(\ref{e0}) as a
function of $x_{0},$ i.e., $p=1/(1+x_{0}^{2}),$ into $g_{+}=0$ of Eq.(\ref
{j0}), \textbf{\ }yielding $x=$ $X(x_{0},t)=x_{0}+t/(1+x_{0}^{2})$. After
the critical time $t_{c}=\frac{8}{9}\sqrt{3}$ there will be caustics at
points $x_{1}(t)$ and $x_{2}(t);$ for each $x$ between these two points
there will be three values of $p$ on the Lagrangian submanifold which
correspond to three initial positions, $x_{i0}$ $[i=1,2,3]$. Associated with
these three different $p^{\prime }s$ there will be three density
distributions $\rho _{i}(x,t),$ which turn out to be

\[
\rho _{i}(x,t)=\frac{(1+x_{i0})^{2}\rho _{0}(x_{i0})}{(1+x_{i0})^{2}-2tx_{i0}%
}. 
\]
At $t_{c}$ the ``flow'' splits at the first caustic point $x_{c}$ into three
partial flows, and thereafter there are two moving infinite ``density
waves'' at the caustic positions.

The singularities at $x_{1}(t)$ and $x_{2}(t)$ are folds, while the
singularity at the ``trifurcation'' point $x_{c},$ (the point at the
critical time where the caustic first begins) is an unstable one if
considered as belonging to the Lagrangian projection at fixed $t$, while it
is stable as a singularity of the family of maps with variable $t,$ called a
metamorphosis (perestroikas).

We mention in passing that Eq.(\ref{j4}) can alternatively be interpreted as
defining a Lagrangian submanifold in the $(x,t;p,-E)$ phase space over the $%
(x,t)$ space-time as base. In that interpretation the fold curves $x_{i}(t)$
meet at the (stable) cusp point $(x_{c},t_{c})$ where the caustic begins.

The ideas described here can (in principle) be extended to H-J theory with
arbitrary Hamiltonians. From a complete solution of the H-J equation on an $%
n $-dimensional configuration space (i.e., one that depends on $n$
independent constants) it is possible to construct a solution $S(x^{a},t)$
from arbitrary Cauchy data, $S_{0}(x^{a})$ and study the evolving Lagrangian
submanifold with the development of the critical points and density waves.
In fact the use of this idea has been proposed and extensively developed in
order to account for the origin of large scale structure in the early
universe\cite{Z,ASZ,SZ}. In Secs.VI and VII in the discussion of generating
families, we will return to this issue.

\section{Multiple Uses for the Eikonal}

Another interesting and useful application of the ideas of Sec.II is to wave
propagation in arbitrary space-times. The treatment is essentially
kinematic, the dynamics enters in the fact that we are assuming that we can
solve the eikonal equation and that we can produce families of solutions at
will. For simplicity of presentation, we will take a rather unsophisticated
approach to the solution of algebraic equations, assuming that almost always
a solution exists. Later, in Section VI, we will give a more sophisticated
treatment of the same issues

We begin with an arbitrary four dimensional manifold, $M^{4}$ , with a
Lorentzian metric $g$ given in some local coordinate system $x^{a}$, by $%
g_{ab}(x$). We want to consider the null hypersurfaces of $g$ in $M^{4}$;
e.g., the hypersurfaces of constant phase in the geometric optics (high
frequency) limit of the Maxwell equations on $M^{4}$. These hypersurfaces
can be described as the level surfaces of functions $S=S(x^{a}$) satisfying
the eikonal equation, (or massless H-J equation) namely 
\begin{equation}
g^{ab}\partial _{a}S\partial _{b}S=0.  \label{9}
\end{equation}

A solution $S=S(x^{a})$ will be referred to as an eikonal. Though later we
will discuss (in a special case) the problem of generating solutions to Eq.(%
\ref{9}), at the start we will assume that we have been given a solution $%
S(x^{a})$ that is continuous but perhaps only piece-wise smooth. The level
``surfaces'' of $S$ might have self-intersections and have sharp edges, as
for example in Figs.3 and 4.

We will consider several different uses for the $S(x^{a}).$

1. First we write $S(x^{a})=S(x^{A},r,t)$, A = 1, 2, having made an
arbitrary decomposition of the four space into a one parameter family of
three dimensional spaces $(t=constant)$; these three surfaces are in turn
foliated by families of two dimensional surfaces, $M^{2}$, with $x^{A}$ as
coordinates. These 2-surfaces are parametrized by $r$ on each $t=constant$
three-surface. (One might think of $t=constant$, as space-like surfaces,
with $r$ as a radial coordinate and the $x^{A}$ as the local angular
coordinates on the two-surface, though there are many alternate pictures one
could make.) We now consider the cotangent bundle over each $M^{2}$, with $%
\omega =dx^{A}\wedge dp_{A}$ and construct a Lagrangian submanifold, $L^{2},$
on it in the following manner. From $S=S(x^{A},r,t)$, fix $S=$ $s_{0}$ and $%
t=$ $t_{0}$ and solve for $r$, obtaining

\begin{equation}
r=R(x^{A},t_{0},s_{0}).  \label{10}
\end{equation}

Using this as a generating function for $L^{2}$, we have that

\begin{eqnarray*}
p_{A} &=&\partial _{A}R(x^{A},t_{0},s_{0}), \\
x^{A} &=&x^{A}
\end{eqnarray*}
defines a Lagrangian submanifold.

2. Any one of these symplectic manifolds, $M_{S}^{2},$ can now be
contactified by adding the coordinate $r$. The contact form is then $\alpha
=dr-p_{A}dx^{A}.$

A Legendrian submanifold $E^{2}$ in $M_{S}^{2}\times \Bbb{R}$ is defined by

\begin{eqnarray}
p_{A} &=&\partial _{A}R(x^{A},t_{0},s_{0}),  \label{11} \\
\text{ }x^{A} &=&x^{A},  \nonumber \\
r &=&R(x^{A},t_{0},s_{0}),  \nonumber
\end{eqnarray}
with the projection to the $(x^{A},r)$ space (i.e., the three dimensional
space $t=t_{0})$ given by $r=R(x^{A},t_{0},s_{0}).$ This describes a wave
front constructed from the intersection of the null ``surface'', $%
s_{0}=S(x^{a})$ with the $t=$ $t_{0}$ three-surface. Note that by the
assumptions in this construction the wavefronts will not have any
singularities; however if evolved to later times they in general do develop
singularities.

3. An alternate way of looking at this evolution is to go back to $S(x^{a})$
and view it as $S(x^{i},t),$ with $x^{i}=(x^{A},r),$ and again take $S=s_{0}$
and then solve for $t=T(x^{i},s_{0})$. Now we treat a manifold (a time
slice), $M^{3}$, with coordinates $x^{i}$, as the base space of a symplectic
bundle, $(x^{i},p_{i})$ with form

\[
\omega =dx^{i}\wedge dp_{i}. 
\]

A Lagrangian submanifold can be obtained from the generating function, $%
T(x^{i},s_{0}),$ with projection to $M^{3}$. Again the construction used
here precludes caustics; however generalizations to be considered later do
lead to caustics which consists of the singularities of the evolving wave
fronts.

\begin{remark}
: Note that there is a completely different Lagrangian submanifold in the
same, $(x^{i},p_{i}),$ symplectic space also constructed from $S(x^{a})$; it
arises from allowing the value of $S$ to vary but keeping $t_{0}$ constant.
A generating function is $S=S(x^{i},t_{0}).$ Its projection (and the
associated caustics) to $M^{3}$ are completely different from those of the
generating function $t=T(x^{i},s_{0})$. They are the caustics associated
with families of null surfaces studied at one instant of time; contrasted
against the previous case of the projection of one null surface $S=S_{0}$ in
space-time into the three space of the $x^{i}$ . This distinction often is
the source of considerable confusion.
\end{remark}

4. This Lagrangian structure, obtained from $T(x^{i},s_{0})$, can be
contactified by adding to $M^{3}$ the coordinate $t$, with contact form $
\alpha =dt-p_{i}dx^{i}.$ Now taking the generating function as $%
t=T(x^{i},s_{0}),$ we obtain a Legendrian submanifold of ($x^{i},p_{i},t)-$
space. Its projection to space-time, $M^{4}=(x^{a})=(x^{i},t)$ is a null
``three-surface'', referred to as a ``big wave front'' by Arnold. This is
the same ``surface'' as described by $s_{0}=S(x^{a});$ a level surface of
the eikonal.

5. As the last use of the eikonal, we mention that taking $S=S(x^{a})$ as
the generating function for a Lagrangian submanifold in the symplectic
manifold over $M^{4}$ given by $p_{a}$ = $\partial _{a}S(x^{a})$ there will,
in general, be three dimensional caustics associated with its projection to $%
M^{4}.$ We do not know of a geometric use for this construction.

$\spadesuit $Often in a physical discussion one is interested in a steady
state situation where a light source (say a point or a two-surface) would
light up and remain on as a source (in time) of families of wave fronts.
\{We assume in this discussion that the metric (or a conformally related
metric) in the eikonal equation does not depend on time.\} The families of
wave fronts would look exactly alike at every instant of time $t$. The
problem is to solve the eikonal equation so that the initial or Cauchy data, 
$S(x^{i},t_{0})=S_{0}(x^{i}),$ corresponds to the evolution of one wave
front obtained from its normal evolution from the given source surface, that
is projected back to the $t=t_{0}$ three-surface. This can be accomplished
by returning to item \#3, where the evolving wave fronts on the three
manifold $M^{3}$ of the ($x^{i})$, was described by the level 2-surfaces of $%
t=T(x^{i},s_{0})$. Treating the two-surface defined by $t_{0}=T(x^{i},s_{0})$
as the source surface and ignoring $s_{0}$ since it is a given constant, we
can define the Cauchy data by $S_{0}=T(x^{i})$; the level surfaces of this
function $S_{0}$ are the wave fronts at $t=t_{0}$ of light emitted by the
source at earlier times. This is the situation that arises in the discussion
of gravitational lensing; it is assumed that there is a fixed source in a
(conformally stationary) space-time that continuously emits light.

A closely associated point of view to this is to consider the ``\textit{%
time-independent}'' eikonal equation (defined only in conformally stationary
space-times), namely

\begin{equation}
g^{ij}\partial _{i}\widehat{T}\partial _{j}\widehat{T}-2g^{i0}\partial _{i}%
\widehat{T}+g^{00}=0.  \label{40}
\end{equation}

This equation can be obtained by substituting the ansatz, $S=t-\widehat{T}%
(x^{i})$ into the original eikonal equation. This equation is satisfied by
the Cauchy data $S_{0}=T(x^{i})$ of the previous paragraph if we take $T$,
defined implicitly by 
\begin{equation}
s_{0}=S(x^{i},T)  \label{39}
\end{equation}
where $S(x^{i},t)$ satisfies the eikonal equation, $g^{ab}\partial _{a}S%
\partial _{b}S=0.$ Indeed, differentiating Eq.(\ref{39}) with respect to $%
x^{i},$ we obtain $0=$ $\partial _{i}S+(\partial S/\partial t)\partial _{i}T$
or

\qquad \qquad \qquad 
\[
\partial _{i}T=-\frac{\partial _{i}S}{\partial _{t}S} 
\]
which when substituted into Eq.(\ref{40}) leads back to the eikonal equation$%
.$ Thus our ``stationary'' Cauchy data is a \textit{solution} of the\textit{%
\ time-independent} eikonal equation.

In the case of static space-times where the $g^{i0}=0,$ Eq.(\ref{40}) is
often written in terms of the Hamiltonian $H=g^{ij}p_{i}p_{j}-n^{2}(x^{i})=0 
$ with $n^{2}=-g^{00},$ reinterpreted as an effective index of refraction
and $p_{i}=\partial _{i}T.$ This point of view leads to Fermat's Principle
of least time.

$\blacklozenge $ We give a powerful example of how the eikonal equation can
be solved with given Cauchy data. We will assume that a three parameter
family of solutions of the eikonal equation is known. In principle there
always exists such a three parameter family of solutions, though in practice
it is generally very hard to find them exactly. Perturbation techniques
might be needed to approximate them. Nevertheless for the general discussion
we will assume that there exists a solution $S^{*}$ of the eikonal equation
that depends on three independent parameters, i.e., $\qquad \qquad $%
\begin{equation}
S^{*}=S^{*}(x^{i},t,\alpha _{i}).
\end{equation}
This is called a complete integral. We show that a ``general integral''
(which involves an arbitrary function of three variables) can be constructed
from the complete integral in the following fashion: we first add to it an
arbitrary function of the $\alpha _{i}$ , i.e., we consider

\begin{equation}
S^{**}=S^{*}(x^{i},t,\alpha _{i})-F(\alpha _{i})  \label{12}
\end{equation}
which trivially still satisfies the eikonal equation.

We next form the equations

\begin{equation}
\partial S^{**}/\partial \alpha _{i}=\partial S^{*}/\partial \alpha
_{i}-\partial F/\partial \alpha _{i}=0.  \label{12a}
\end{equation}
For the present we assume that it has a solution of the form $\alpha
_{i}=A_{i}(x^{i},t)$. (This is an example of our unsophisticated treatment.
Tacitly we are referring to the implicit function theorem, assuming that the
determinant of $\partial ^{2}S^{**}/\partial \alpha _{i}\partial \alpha _{j}$
is different from zero. We will return later, in Secs.VI and VII, to the
issue of solving Eq.(\ref{12a}) when the determinant vanishes.)

Finally, via $\alpha _{i}=A_{i}(x^{i},t)$, the $\alpha _{i}$ are eliminated
in $S^{**}$ yielding

\begin{equation}
S(x^{i},t)=S^{*}(x^{i},t,A_{i}(x^{i},t))-F(A_{i}(x^{i},t)).  \label{00}
\end{equation}

It is not difficult to show that this $S$ satisfies the eikonal equation\cite
{37}. The $x^{a}$ derivatives of $S(x^{i},t)$ involve both the explicit $%
x^{a}$ dependence and the dependence via the $A_{i}(x^{i},t);$ the latter
dependence however drops out because of Eq.(\ref{12a}). Since the eikonal
equation is satisfied as far as the explicit dependence then so does Eq.(\ref
{00}). This solution now depends on an arbitrary function of three
variables, namely $F(\alpha _{i})$.

The task is now to determine $F(\alpha _{i})$ in terms of Cauchy data, $%
S{}_{0}$($x^{i}$). This is accomplished as follows; consider the function $%
\frak{s}(x^{i},\alpha _{i})\equiv S^{*}(x^{i},t_{0},\alpha _{i})-S{}_{0}$($%
x^{i}$) and then construct the three equations

\[
\frac{\partial \frak{s}(x^{j},\alpha _{j})}{\partial x^{i}}\equiv \frac{
\partial (S^{*}-S{}_{0})}{\partial x^{i}}=0. 
\]
Because $S^{*}$ is a complete solution, they can be solved for

\[
\alpha _{i}=A_{i}(x^{j}). 
\]
We now assume that the Cauchy data was chosen so that the last equation can
be algebraically inverted, i.e., 
\[
x^{i}=X^{i}(\alpha _{i}). 
\]
At $t=t_{0}$ we have that 
\begin{equation}
S_{0}(x^{i})=S^{*}(x^{i},t_{0},A_{i}(x^{i}))-F(A_{i}(x^{i})).  \label{13}
\end{equation}
Replacing all the $A_{i}\ $by $\alpha _{i}$ and all the $x^{i}$ by $%
X^{i}(\alpha _{j})$, we have that

\begin{equation}
F(\alpha _{i})=S^{*}(X^{i}(\alpha _{i},t_{0}),\alpha
_{i})-S_{0}(X^{i}(\alpha _{i})),  \label{14}
\end{equation}
i.e., the free $F(\alpha _{i})$ is now expressed in terms of the free Cauchy
data, $S_{0}(x^{i})$ and the complete solution.

\{The construction described here for the solution of the Eikonal equation
in terms of Cauchy data can be easily extended to the H-J equation. See
Sec.VII.\}

$\spadesuit $In Minkowski space, the plane waves provide a complete integral;

\[
S^{*}=x^{i}\alpha _{i}-t\sqrt{(\alpha _{i})^{2}}. 
\]
This allows us to find (in principle - modulo algebraic inversions) all
solutions of the flat-space Eikonal equation with arbitrary Cauchy data. $%
\spadesuit $

\section{A Caveat}

As we pointed out in examples, the principal strength and importance of the
theory of Lagrangian and Legendrian projections lies in its ability to treat
places where the projections are not diffeomorphisms.

Though in many cases it is possible and perhaps even intuitively useful to
treat the projections as being almost always diffeomorphisms with lower
dimensional regions as the exception. One could try to handle the
singularities formally by allowing the generating functions to be
multivalued and only piece-wise smooth and then approach the critical points
as limits of regular ones. However this creates difficulties: the
projections to the base space near critical points are difficult to treat
and the structure of the caustics or wave front singularities are often hard
to ``see''. Even when possible, this approach certainly is inelegant and
ill-defined mathematically. One might have hoped that the cases with
critical points would be exceptions but the opposite is true; the existence
of critical points is generic and one must be able to construct the proper
type of generating function. In Sec. IV where we dealt with solutions to the
eikonal equation we always tacitly assumed that the relevant equations could
be solved for certain specific variables. This was true for most regions but
not in the regions where certain Jacobians vanished and where the critical
points existed. In fact it is the existence of the critical points that was
the obstruction to solving the algebraic equations.

An important question then is; does there exist some general procedure
applicable to physical problems for the construction of Lagrangian or
Legendrian submanifolds with the associated projection maps - including
singularities - so that the issue of finding appropriate choices for the
generating function does not arise and where the associated projection maps
are given in some natural systematic manner. We saw, in examples $\spadesuit
c$ and $\spadesuit f$ how to obtain the Lagrangian and Legendrian
submanifolds (with critical points) without a generating function and then
in Sec.III, discussing the free particle H-J equation, the evolution of the
H-J equation itself suggested the Legendre transformation to a proper
single-valued generating function. We will see in the next section that
there indeed is a systematic method based on the concept of ``generating
families''. We will see that it allows us to construct general Lagrangian
and Legendrian submanifolds associated with either H-J or eikonal evolution
based on arbitrary Cauchy data. However if wanted or needed, it is possible
to construct generating functions, which are local objects, from generating
families which serve to define the submanifolds globally.

\section{Generating Families}

There is a remarkably beautiful method \cite{30,39} for the construction of
single valued (local) generating functions - easily applied in many physical
situations - using what are called ``\textit{generating \ families}''.
Actually one can bypass the generating function construction and go
directly, via the generating \textit{\ families,} to the Lagrangian and
Legendrian submanifolds and associated maps. [Though in the literature of
catastrophe theory what we are calling generating \textit{families} are
frequently called generating functions, we will stay with the terminology
adopted by Arnold.]

We first outline the mathematical ideas behind ``generating families'' and
then show how it can be applied to various physical situations or problems.

We give two methods for the construction of generating families; the first
begins with a generating \textit{function}, a special class of generating 
\textit{families} being constructed from it. Second, from certain
observations concerning the first construction, the procedure can be
generalized to what becomes the full theory of generating families.

For the first method we start with the cotangent bundle $T^{*}M$ equipped
with canonical coordinates $(q^{a},p_{a}).$ For step one, we assume a
Lagrangian submanifold $L$ to be given, with a point $\xi $ on $L$ and a
generating function $G(q^{A},p_{J})$ near $\xi .$ The local embedding of $L$
into $T^{*}M$ is then given by

\begin{equation}
q^{J}=-\partial ^{J}G,\qquad p_{A}=\partial _{A}G.  \label{1x}
\end{equation}
We now define a function $F$ in a neighborhood $U$ of $\xi $ in $T^{*}M$ by

\begin{equation}
F(q^{a},p_{a})=G(q^{A},p_{J})+q^{J}p_{J}.  \label{2x}
\end{equation}
Since $F$ does not depend on $p_{A}$ nor\textit{\ }$G$\textit{\ }on $q^{J},$
it is trivially seen that $F$ identically satisfies the equations

\begin{equation}
\partial ^{A}F=0,\qquad \partial _{J}F=p_{J}.  \label{3x}
\end{equation}

The remaining equations

\begin{equation}
\partial ^{J}F=0,\text{ }\qquad \partial _{A}F=p_{A},  \label{4xz}
\end{equation}
which are\textbf{\ }equivalent to Eqs.(\ref{1x}), define the Lagrangian
submanifold.

The construction of $F$ from $G$ implies that

\begin{equation}
rank(\partial ^{a}F=0,\text{ }\partial _{a}F-p_{a})=n,  \label{6x}
\end{equation}
since the $2n$ equations in Eq.(\ref{6x}) have, by construction, the unique
solution (\ref{1x}).

We henceforth want to forget the $G$ from which $F$ was constructed and
claim: If a function $F(q^{a},p_{a})$ satisfies the rank condition Eq.(\ref
{6x})\textbf{, }then the $2n$ equations

\begin{equation}
\partial ^{a}F=0,\qquad \partial _{a}F=p_{a}  \label{5x}
\end{equation}
can be solved for some $n$ of the set $(q^{a},p_{a}).$ This uniquely yields
an embedded Lagrangian submanifold. Indeed, according to the implicit
function theorem, one can then express $n$ of the variables $(q^{a},p_{a})$
- say $(q^{J},p_{A})$ - in terms of the remaining ones; in other words Eqs.%
\textbf{(}\ref{5x}\textbf{) }implies that

\[
q^{J}=Q^{J}(q^{A},p_{J}),\quad p_{A}=P_{A}(q^{A},p_{J}) 
\]
hold on the $n$-manifold $L$ with (local) coordinates\textbf{\ }$%
(q^{A},p_{J}).$ If the identity

\[
dF=\partial _{a}Fdq^{a}+\partial ^{a}Fdp_{a} 
\]
on $T^{*}M$ is pulled back to $L$, then because of Eqs.\textbf{(}\ref{5x}%
\textbf{), }one has the result that: on $L$

\begin{equation}
dF=p_{a}dq^{a}=\partial _{a}Fdq^{a}=\kappa ,  \label{6xz}
\end{equation}
i.e., the restriction of $F$ to $L$ is a potential for the one-form $\kappa $
on $L$ and hence $\omega =0$. Thus $L$ is, in fact, Lagrangian. Moreover, on 
$L,$ we see that $F-q^{J}p_{J}\equiv G$ obeys Eqs.\textbf{(}\ref{1x}\textbf{%
).}

This formulation in terms of $F,$ - i.e., $F$ being any function obeying the
rank condition, Eq.(\ref{6x}), - allows the construction of Lagrangian
submanifolds which may have regular points as well as critical points and
which might require, for their local descriptions, several different
generating functions. In this sense, the description in terms of $F$ via
Eqs.(\ref{6x}) and (\ref{5x}) is more general and ``less local'' than the
one in terms of $G$ and Eq.(\ref{1x}). Any Lagrangian submanifold can
locally be obtained from some $F.$

The foregoing argument generalizes immediately to the Legendre case. We
simply add to $F$, (which is a function on $T^{*}M)$ the additional variable 
$u$ $\epsilon $ $\Bbb{R},$ and form $u=F$. Then a \thinspace Legendrian
submanifold, $E,$ on $T^{*}M\times \Bbb{R}$ is given by

\begin{equation}
u=F,\quad \text{ }p_{a}=\partial _{a}F,\text{ }\quad \partial ^{a}F=0.
\label{7x}
\end{equation}

The complete theory of generating families now arises as a generalization of
the preceding construction. At first sight there appear to be substantial
differences but on closer observation we see that it really is a
generalization. We will show later how the previous case is a specialization
of the general theory.

The basic idea is to start with a configuration space, $M^{n}$, of dimension 
$n$ and then enlarge it to a space $M^{n+m}=M^{n}\times M^{m}$ of dimension $%
n+m$ , with local coordinates ($q^{a}$, $s^{J}$), $a$ $=1,....,n$ and $J$ = $%
1,...,m$. The dimensions $n$ and $m$ are arbitrary. A ``generating
function'' $\frak{F}(q^{a},s^{J})$ defined on the large space, $M^{n+m}$ ,
is then chosen, e.g., by physical or geometric arguments (examples of which
will be given shortly).

$\frak{F}(q^{a},s^{J})$ is arbitrary except for the following rank
condition: The equations

\begin{equation}
\frac{\partial \frak{F}}{\partial s^{J}}\equiv \partial _{J}\frak{F}=0
\label{Bx}
\end{equation}
admit solutions for some $m$ of the set ($q^{a},s^{J}$), and whenever they
hold, the $(n+m)\times m$ matrix

\begin{equation}
\lbrack \frak{F}_{Ja}\equiv \frac{\partial ^{2}\frak{F}}{\partial
s^{J}\partial q^{a}},\quad \frak{F}_{JK}\equiv \frac{\partial ^{2}\frak{F}}{
\partial s^{J}\partial s^{K}}]  \label{B}
\end{equation}
has rank $m.$

From $\frak{F}(x^{a},s^{J})$, by an ingenious method, one can then either
construct appropriate generating \textit{functions} on the cotangent bundle
over $M^{n}$ and hence a Lagrangian submanifold or instead, directly
construct the Lagrangian submanifold from the generating \textit{family}.

Since the considerations are essentially local we can consider $M^{n+m}=$ $%
R^{n}$x $R^{m}.$

We first state the main result; namely how to construct an $n$-dimensional
(Lagrangian) submanifold from the generating family $\frak{F}(x^{a},s^{J}).$
This is then followed by the proof that the submanifold so constructed is in
fact Lagrangian.

We first use the function $\frak{F}(x^{a},s^{J})$ as a generating function
to generate a Lagrangian section $\widehat{L}$ in the cotangent space over $%
M^{n+m},$

\begin{eqnarray}
p_{a} &=&\partial \frak{F}/\partial q^{a},\text{ \quad }q^{a}=q^{a}
\label{AA} \\
\Pi _{J} &=&\partial \frak{F}/\partial s^{J},\text{ \quad }s^{J}=s^{J}. 
\nonumber
\end{eqnarray}

We then define a subset of $M^{n+m}$ by imposing the extremal condition

\begin{equation}
\Pi _{J}=\partial \frak{F}/\partial s^{J}(q^{a},s^{K})=0.  \label{C}
\end{equation}

According to the rank condition, the solutions of this equation form an $n$
-dimensional submanifold of $M^{n+m},$ that can be expressed by

\begin{equation}
q^{a}=Q^{a}(y^{b}),\qquad s^{J}=S^{J}(y^{b})  \label{10xz}
\end{equation}
When these are substituted into $p_{a}=\partial _{a}\frak{F,}$ one obtains $%
p_{a}=P_{a}(y^{b}):$ the equations

\begin{equation}
q^{a}=Q^{a}(y^{b}),\quad s^{J}=S^{J}(y^{b}),\quad p_{a}=P_{a}(y^{b}),\quad
\Pi _{J}=0  \label{10xzz}
\end{equation}
define an $n$-dimensional submanifold $N$ of the large phase space. By its
construction, $N$ is the intersection of $\widehat{L}$ and the submanifold $%
P $ of $T^{*}M^{n+m}$ defined by $\Pi _{J}=0$.

The submanifold of $T^{*}M^{n}$ defined by

\begin{equation}
q^{a}=Q^{a}(y^{b}),\quad \quad p_{a}=P_{a}(y^{b})  \label{10xzzz}
\end{equation}
is Lagrangian.

What follows is the proof of this contention. As the proof is rather
technical and difficult the reader might simply prefer to accept the
contention and bypass the proof. Doing so does not greatly affect the
understanding of, or the ability to use, generating families. The proof is
given for completeness.

Proof: Let $\xi $ be a point of $N.$ The dimensions of the tangent spaces $%
N_{\xi },\widehat{L}_{\xi },P_{\xi },$ are $n,n+m,2n+m,$ respectively. Since 
$(n+m)$ independent vectors at $\widehat{L}_{\xi },$(obtained from the
derivatives of Eq.(\ref{AA})$)$ using $\Pi _{J}=\partial _{J}\frak{F}$ and $%
p_{a}=\partial _{a}\frak{F})$ have the form

\[
V_{(a)}=(\partial _{ab}\frak{F)}\partial ^{b}+\frak{F}_{aK}\text{ }\partial
^{K}+\partial _{a},\text{ }\quad V_{(J)}=\frak{F}_{bJ}\partial ^{b}+\frak{F}
_{JK}\partial ^{K}+\partial _{J} 
\]
and vectors at $P_{\xi }$ have the form

\[
Y=Y^{a}\partial _{a}+Y^{J}\partial _{J}+Y_{a}\partial ^{a},
\]
(with $Y^{a},Y^{J},Y_{a}$ arbitrary and $Y_{J}=0$) we see immediately that $%
\widehat{L}_{\xi }+P_{\xi }$ spans the tangent space of $T^{*}M^{n+m}$ at $%
\xi $ and hence dim$(\widehat{L}_{\xi }+P_{\xi })=2n+2m;$ i.e., $\widehat{L}%
\ $ and $P$ intersect transversely. (This statement is the geometric
reformulation of the rank condition.)

The critical point to be established next is that the projection of $N$ into
the small phase space $T^{*}M^{n},$ a projection along the $s^{J}$-direction
is everywhere a local diffeomorphism, so that the image $L$ is an\textit{\ }$%
n$-dimensional submanifold of $T^{*}M^{n},$ given by, Eq.(\ref{10xzzz}),

\[
q^{a}=Q^{a}(y^{b}),\quad p_{a}=P_{a}(y^{b}). 
\]
To prove that, one has to show that no (non-vanishing) vector tangent to $N$
is annihilated by the projection. Following Arnold, this can be done
elegantly as follows.

We first note that the kernel of the projection consists of all vectors of
the form $X=X^{J}\partial _{J}$ (i.e., vectors in the $s^{J}$-directions)
and then observe that, (from the skew-orthogonal product of tangent vector
of $T^{*}M^{n+m},$ defined by $[X,Y]\equiv
X^{J}Y_{J}-X_{J}Y^{J}+X^{a}Y_{a}-X_{a}Y^{a}),$ the kernel is skew-orthogonal
to all the vectors $Y$ tangent to $P,$ i.e., from $Y=Y^{a}\partial
_{a}+Y^{J}\partial _{J}+Y_{a}\partial ^{a},$ $[X,Y]=0.$ Suppose now that $X$
is in the kernel and tangent to $N$. Then $X$ is also tangent to $\widehat{L}
$ since $N $ $\subset \widehat{L}$. Therefore $X$ is skew-orthogonal to both 
$P$ and $\widehat{L}$ (since $\widehat{L}$ is Lagrangian all vectors in $%
\widehat{L}_{\xi }$ are skew-orthogonal). But since the tangents to $P$ and $%
\widehat{L} $ together span the total tangent space of $T^{*}M^{n+m}$ -
i.e., transversality - $X$ is skew-orthogonal to ``everything'', and thus $%
X=0$, which was to be shown.

The submanifold $L$ given by Eqs.(\ref{10xzzz})\textbf{\ }is, in fact,
Lagrangian. This again follows by pulling back to $L$ the identity

\[
d\frak{F}=\partial _{a}\frak{F}dq^{a}+\partial _{J}\frak{F}ds^{J} 
\]
which results in

\begin{equation}
d\frak{F}=p_{a}dq^{a}=\kappa .\qquad \qquad \text{QED}  \label{11x}
\end{equation}

Note that any Lagrangian submanifold of $T^{*}M^{n}$ can be obtained by the
foregoing construction. Suppose that $L$ is given locally by $K(q^{A},p_{J})$
as in Eq.(\ref{22*})\textbf{. }Then the generating family (of the type
considered in Eq.(\ref{2x})).

\[
\frak{F}(q^{a},p_{J})=K(q^{A},p_{J})+q^{J}p_{J}, 
\]
(with $p_{J}=s^{J}),$ reproduces $L$, as is easily verified.

The projection of $L$ to the base is, of course given by

\[
q^{a}=Q^{a}(y^{b}). 
\]
Taking into account how $Q^{a}(y^{b})$ was obtained via Eq.(\ref{C}), one
can see that the kernel of that projection is determined by the solution $%
X^{K}$ of

\[
\frak{F}_{JK}X^{K}=0, 
\]
thus the critical points of $L$ are given by 
\begin{equation}
D\equiv \left| \frac{\partial ^{2}\frak{F}}{\partial s^{J}\partial s^{K}}
\right| =0.  \label{D}
\end{equation}

We may summarize and geometrically interpret the preceding construction as
follows: For each fixed $s^{J},$ $p_{a}=$ $\partial _{a}\frak{F}(q^{a},s^{J})
$ defines a singularity-free Lagrangian submanifold of $T^{*}M^{n},$ i.e., $%
\frak{F,}$ acting as a generating function, defines an $m$-parameter family
of ``regular'' Lagrangian submanifolds. By solving $\partial _{J}\frak{F}=0,$
i.e., Eq.(\ref{C}), and inserting them into $p_{a}=$ $\partial _{a}\frak{F}%
(q^{a},s^{J}),$ we obtain $p_{a}=$ $P_{a}(y^{b}),$ which with $%
q^{a}=Q^{a}(y^{b}),$ provides another Lagrangian submanifold, the \textit{%
envelope} of the former family. This Lagrangian submanifold has the
projection map $\pi :y^{b}\rightarrow q^{a}=Q^{a}(y^{b}).$ Its critical
points are given as those points where the rank of the Jacobian matrix $%
\partial _{a}Q^{b}$ is not maximal or equivalently, where Eq.(\ref{D} )%
\textbf{\ }holds.

Now we can also see that the previous construction via Eq.(\ref{5x}) is
included in the general case. If $m=n$ and if the first matrix in Eq.(\ref{B}
), i.e., $\frak{F}_{Ka},$ has rank $m,$ then one can express the $s^{J}$ as
functions of the $p_{a}$, and $F$($q^{a},p_{a}$) $\equiv \frak{F(}
q^{a},s^{J}(p_{a})\frak{)}$ is a generating family of the former kind.

Eq.(\ref{10xzzz}) represents the Lagrangian submanifold in terms of some
coordinates $y^{b}.$ Due to the implicit function theorem, the $y^{b}$ can
always (locally) be chosen as subsets of the ($q^{a},s^{J}$).

We now consider the possible cases:

\#1. Let us first assume that at a solution point, ($q_{0}^{a},s_{0}^{J}),$

\begin{equation}
D\equiv \left| \frak{F}_{JK}\right| \neq 0.
\end{equation}
Then, Eqs.(\ref{C}) can be solved uniquely for all the $s^{J},$

\begin{equation}
s^{J}=S^{J}(q^{a}).  \label{E}
\end{equation}
This result can be inserted into $\frak{F}(q^{a},s^{J})$ so that

\begin{equation}
\frak{F}(q^{a},s^{J})\Rightarrow G(q^{a})=\frak{F}(q^{a},S^{J}(q^{a}))
\label{EE}
\end{equation}
yields a generating function $G(q^{a})$ for a Lagrangian submanifold. From
the general theory 
\begin{equation}
p_{a}=\partial G/\partial q^{a},\text{ }q^{a}=q^{a}
\end{equation}

\noindent with a trivial (diffeomorphism) Lagrangian map.

Conversely when $D=$ $0,$ at ($q_{0}^{a},s_{0}^{J}),$ the Lagrangian
projection is not a diffeomorphism in any neighborhood of the point, i.e.,
we have a Lagrangian singularity there as noted in connection with Eq.(\ref
{D}). The vanishing of $D$ is thus the necessary and sufficient condition
for the occurrence of a caustic at the point in question.\qquad

\#2. The other case to consider is when the $m$ equations $\partial _{J}%
\frak{F}=0$ can be algebraically solved for a mixture of some $q^{a}$'s and
some $s^{J}$'s, i.e., where the solutions have the form

\begin{equation}
q^{a^{\prime }}=Q^{a^{\prime }}(q^{a^{\prime \prime }},s^{K^{\prime \prime
}}),\quad s^{J^{\prime }}=S^{J^{\prime }}(q^{a^{\prime \prime
}},s^{K^{\prime \prime }}),  \label{EEE}
\end{equation}
$\qquad \qquad \qquad $with $m$ variables ($q^{a^{\prime }},s^{J^{\prime }})$
and $n$ variables $(q^{a^{\prime \prime }},s^{K^{\prime \prime }})$ such
that at least one $s^{K^{\prime \prime }}$ occurs. (The set of $q^{a^{\prime
\prime }}$ might be empty.) The Lagrangian submanifold, parametrized by the $%
n$ variables $(q^{a^{\prime \prime }},s^{K^{\prime \prime }}),$ is now given
by\qquad

\begin{eqnarray}
q^{a^{\prime }} &=&Q^{a^{\prime }}(q^{a^{\prime \prime }},s^{K^{\prime
\prime }})  \label{I} \\
q^{a^{\prime \prime }} &=&q^{a^{\prime \prime }},  \nonumber \\
p_{a} &=&\partial _{a}\frak{F}=P_{a}(q^{a^{\prime \prime }},s^{K^{\prime
\prime }}).  \nonumber
\end{eqnarray}

The generating function

\begin{equation}
S(q^{a^{\prime \prime }},p_{a^{\prime }})=\frak{F}{\Large (}Q^{a^{\prime }}%
{\small (}q^{a^{\prime \prime }},s^{K^{\prime \prime }}{\small )}
,q^{a^{\prime \prime }},S^{J^{\prime }}(q^{a^{\prime \prime }},s^{K^{\prime
\prime }}{\small )},s^{K^{\prime \prime }}{\large )}-p_{a^{\prime
}}Q^{a^{\prime }}{\small (}q^{a^{\prime \prime }},s^{K^{\prime \prime }}%
{\small ),}  \label{II}
\end{equation}
which does not depend on $s^{K^{\prime \prime }},$ yields the same
submanifold as do Eqs.(\ref{I}). To see that $S,$ in fact, does not depend
on $s^{K^{\prime \prime }},$ one first treats the right side as a function
of ($q^{a^{\prime \prime }},p_{a^{\prime }},s^{K^{\prime \prime }})$ and
then by differentiating with respect to $s^{K^{\prime \prime }},$and using
Eqs.(\ref{I}), one sees that the derivative vanishes.

Since, from generating functions for Lagrangian submanifolds one can
construct a contact coordinate (see Eqs.(\ref{5}) and (\ref{5*})) and hence
a Legendrian submanifold and Legendrian map, the construction of the
Lagrangian submanifolds via generating \textit{families }rather than
generating\textit{\ functions}, is easily extended (see Eq.(\ref{7x})) to
the Legendrian submanifolds and maps via

\[
u=\frak{F(}q^{a},s^{J}),\text{    }\partial _{J}\frak{F=}0,\text{     }p_{a}=%
\partial _{a}\frak{F.}
\]

\section{Applications of Generating Families}

Since many or perhaps most of the applications in physics of Lagrangian and
Legendrian submanifolds and maps are associated with dynamical or optical
systems and appear to be either directly or indirectly associated with
Hamilton-Jacobi theory or the related eikonal equation we will confine our
discussion to showing how generating families can be constructed for
specific physical situations.

$\spadesuit $ We begin with a simple but important physical model. Consider
four dimensional Minkowski space-time foliated by the standard $t=const.,$
space-like three surfaces $\sum_{t}$ $\Leftrightarrow \Bbb{R}^{3}$, with
Cartesian coordinates $x^{i}$. We choose at $t=0$, an arbitrary two surface, 
$\frak{S}$, in $\sum_{0}$ that ``lights-up'', i.e., that is to be a source
of light, with local coordinates ($s^{J}$), $J=1,2$, i.e., $%
x^{i}=x_{0}^{i}(s^{J}),$ which describes $\frak{S}$ parametrically. The $%
(x^{1},x^{2},x^{3})$ in $\Bbb{R}^{3}$ are the points of physical space
(observation points) that will be reached by light rays from $\frak{S}$. At
each instant of time $t$, the light fills a region bounded by two ``small
wavefronts'' - from the ``incoming'' and ``outgoing'' rays. In space-time
these small wavefronts, as time evolves, form a pair of null hypersurfaces
(``big'' wavefronts), whose intersection \textit{is} $\frak{S.}$ We wish to
find these small wavefronts.

Let the function $t=\frak{F}(x^{1},x^{2},x^{3},s^{1},s^{2})$ represent the
time it takes for light to go from $\frak{S}$ to the observation point, $%
x^{i}$. From the constancy of the speed of light, $c=1$, we have that

\begin{equation}
t=\frak{F}(x^{1},x^{2},x^{3},s^{1},s^{2})=\sqrt{
(x^{i}-x_{0}^{i}(s^{J}))(x^{i}-x_{0}^{i}(s^{J}))}  \label{24}
\end{equation}
which we will write as

\[
\frak{F}=\sqrt{(\mathbf{r}-\mathbf{r}_{0}(s^{J}))\cdot (\mathbf{r}-\mathbf{r}
_{0}(s^{J}))}. 
\]

First of all we define, in accordance with Eq.(\ref{AA})\qquad \qquad 
\begin{equation}
\mathbf{p=}\frac{\partial \frak{F}}{\partial \mathbf{r}}=\frac{(\mathbf{r}-%
\mathbf{\ r}_{0}(s^{J}))}{\left| \mathbf{r}-\mathbf{\ r}_{0}(s^{J})\right| }.
\label{24a}
\end{equation}

From $\Pi _{J}=\partial \frak{F}/\partial s^{J}=0$, \textbf{\ }we have that

\begin{equation}
\Pi _{J}=-\frac{(\mathbf{r}-\mathbf{\ r}_{0}(s^{J}))}{\left| \mathbf{r}-%
\mathbf{\ r}_{0}(s^{J})\right| }\cdot \mathbf{T}_{J}=-\mathbf{p\cdot T}_{J}=0
\label{25}
\end{equation}
with $\mathbf{T}_{J}$($s^{K}$) = $\partial \mathbf{r}_{0}$/$\partial s^{J}$,
the two tangent vectors to the surface . (The physical meaning of $\partial 
\frak{F}/\partial s^{J}=0$ is that, since $t=\frak{F}(\mathbf{r,r}%
_{0}(s^{J})),$ the travel time of a ray leaving from the surface at $\mathbf{%
r}_{0}(s^{J})$ and arriving at $\mathbf{r}$ is an extremal (minimum). We
see, below, that to satisfy this condition, rays must leave normal to the
surface, $\frak{S.}$

We can solve the Eqs.(\ref{25}) by introducing the unit normal to $\frak{S}$
, given by $\qquad \qquad \qquad \qquad \qquad \qquad \qquad $%
\begin{equation}
\mathbf{n}(s^{J})=\frac{\mathbf{T}_{1}\times \mathbf{T}_{2}}{\left| \mathbf{T%
}_{1}\right| \left| \mathbf{T}_{2}\right| }  \label{25a}
\end{equation}

\noindent and using the fact that Eq.(\ref{25}) implies that

\begin{equation}
\text{ }\mathbf{r}=\mathbf{r}_{0}(s^{J})\mathbf{+}\text{ }v\mathbf{n}(s^{J}).
\label{25b}
\end{equation}
Thus from Eq.(\ref{24a}),

\begin{equation}
\mathbf{p}=\mathbf{n}(s^{J})  \label{25c}
\end{equation}
i.e., if Eq.(\ref{25}) holds, the momentum is the unit normal to the surface 
$\frak{S.}$

Eq.(\ref{25b}), for each fixed value of $v$, defines a small wavefront with
the two signs of $v$ yielding the incoming and outgoing fronts. For
sufficiently large \TEXTsymbol{\vert}$v|$, these fronts could develop
singularities. For examples, see Figs. 6 and 7.

Eqs.(\ref{25b}) and (\ref{25c}) define a (three dimensional) Lagrangian
submanifold in the six dimensional phase space of ($\mathbf{r,p),}$ in terms
of the parameters $v$ and $s^{J},$ while the Lagrangian map $\pi $ is given
by Eqs.(\ref{25b}).

Now with the use of generating families, this example generalizes (from $2$
to $3$ dimensional configuration spaces), the construction $\spadesuit c$
from Sec.II.

The extension of this construction to a Legendrian submanifold, $E,$
consists of simply adding $t$ as the contact coordinate and using $t=v$ with
Eqs.(\ref{25b}) and (\ref{25c}) to define $E$, i.e.,

\begin{eqnarray}
\mathbf{r} &=&\mathbf{r}_{0}(s^{J})\mathbf{+}v\mathbf{n}(s^{J}), \\
\mathbf{p} &=&\mathbf{n}(s^{J}), \\
t &=&v,
\end{eqnarray}
while the projection, the Legendrian map, to the space-time, ($\mathbf{r,}%
t), $ is given by

\begin{eqnarray}
\mathbf{r} &=&\mathbf{r}_{0}(s^{J})\mathbf{+}v\mathbf{n}(s^{J}), \\
t &=&v.
\end{eqnarray}
(Compare with $\spadesuit f$ of Sec.II.) Qualitatively these examples can be
generalized to arbitrary four dimensional Lorentzian space-times\cite
{33,HKP,SNG2}.

As was stated earlier the critical points of the Lagrangian map can be
calculated either from the vanishing of the Jacobian of that map or from

\[
D(s^{J},v)\equiv \left| \frac{\partial ^{2}\frak{F}}{\partial s^{J}\partial
s^{K}}\right| =0. 
\]

Directly from the latter expression we have, after a brief calculation, that

\begin{equation}
\frac{\partial ^{2}\frak{F}}{\partial s^{J}\partial s^{K}}%
=v^{-1}(g_{JK}-vh_{JK})  \label{26}
\end{equation}
where $g_{JK}=\mathbf{T}_{J}\cdot \mathbf{T}_{K}$ and $h_{JK}=$ $\mathbf{n}
(s^{J})\cdot \partial \mathbf{T}_{K}/\partial s^{J}$ are respectively the
first and second fundamental forms (or respectively, the induced metric and
extrinsic curvature tensors) of the surface, $\frak{S}$ $\frak{.}$ The
critical points (determined by the vanishing of the determinant $D,$ of Eq.(%
\ref{26})) are then given by the values of $v=\left| \mathbf{r}-\mathbf{r}%
_{0}\right| $ such that\qquad \qquad 
\begin{equation}
v^{2}D=g+v(g_{11}h_{22}-2g_{12}h_{12}+g_{22}h_{11})+v^{2}h=0  \label{27}
\end{equation}
where $g$ and $h$ are the determinants of $g_{JK}$ and $h_{JK}.$ The two
roots $v_{1}(s^{J})$ and $v_{2}(s^{J}),$ of Eq.(\ref{27}), can be recognized
from the differential geometry of $2$-surfaces in $\Bbb{R}^{3}$, as defining
the two principal radii of curvature of $\frak{S,}$ and we have the
classical result that:

\begin{itemize}
\item  The caustic of a two-surface $\frak{S,}$ acting as a light-source,
consists of the loci of the principal centers of curvature of that surface
and are given by 
\begin{eqnarray}
\mathbf{r(}s^{J}) &=&\mathbf{r}_{0}\mathbf{(}s^{J})+\mathbf{n}v_{1}(s^{J}),
\label{27*} \\
\qquad \mathbf{r(}s^{J}) &=&\mathbf{r}_{0}\mathbf{(}s^{J})+\mathbf{n}%
v_{2}(s^{J}).  \label{27**}
\end{eqnarray}
In other words, it consists of two different two-surfaces, each given
parametrically in terms of $s^{J}$ by Eqs.(\ref{27*}) and (\ref{27**}).
These two surfaces touch each other whenever $v_{1}(s^{J})=v_{2}(s^{J});$ in
other words on the normals from \textit{umbilic} points of $\frak{S}$ where
the two radii of curvature coincide\cite{Klingenberg,Blaschke}. On the
caustic point, associated with the umbilic point, there occurs what is
called an ``umbilic'' singularity. Other ``singularities'' of the caustic
surfaces, which are cusp ridges and swallowtails, can be analyzed in terms
of the local differential geometry of the surface $\frak{S.}$ They are
associated with extremals of the curvatures $(k_{1}=v_{1}^{-1},$ $%
k_{2}=v_{2}^{-1}$) on the curves of a principal curvature coordinate system.
\end{itemize}

[An alternative way to obtain Eq.(\ref{27}) is to calculate and set to zero
the Jacobian of Eq.(\ref{25b});

\begin{equation}
J=\left| \frac{\partial \mathbf{r}}{\partial p},\frac{\partial \mathbf{r}}{%
\partial s^{1}},\frac{\partial \mathbf{r}}{\partial s^{2}}\right| =\mathbf{\
n\cdot \{(T}_{1}+v\frac{\partial \mathbf{n}}{\partial s^{1}}\mathbf{)\times
(T}_{2}+v\frac{\partial \mathbf{n}}{\partial s^{2}}\mathbf{=}0.  \label{27a}
\end{equation}
By using $\mathbf{n}$ from Eq.(\ref{25a}) and the identity $(\mathbf{A\times
B})\cdot (\mathbf{C\times D})=(\mathbf{A\cdot C})(\mathbf{B\cdot D})-(%
\mathbf{A\cdot D})(\mathbf{B\cdot C})$ with the definition of the first and
second fundamental form, Eq.(\ref{27a}) is seen to be identical to Eq.(\ref
{27}).]

\begin{remark}
We mention, without entering into the details, that from Eqs.(\ref{II}),(\ref
{24}) and (\ref{25}) one can construct one of several generating functions
for this case. A typical one, valid if $n_{z}\neq 0,$ takes the form 
\begin{equation}
G(z,p_{x},p_{y})=zn_{z}(s^{J})-\mathbf{r}_{0}(s^{J})\cdot \mathbf{n}(s^{J})
\end{equation}
where the $s^{J}$ are given implicitly as functions of the $(p_{x},p_{y})$
by $(p_{x},p_{y})=(n_{x},n_{y}).\spadesuit $
\end{remark}

A much larger class of examples to which generating families can be applied
is given by the following:

$\spadesuit .$ Consider any (autonomous) Hamiltonian system with phase space
coordinates ($q^{a},p_{a})$ and Hamiltonian

\qquad \qquad \qquad \qquad $H=H(q^{a},p_{a}),$ $a=1.....n,$ ($%
H:T^{*}M\rightarrow \Bbb{R)}$

\noindent and associated (H-J) equation\qquad 
\begin{equation}
\frac{\partial S}{\partial t}+H(q^{a},\frac{\partial S}{\partial q^{a}})=0%
\text{\quad (}S:M\times \Bbb{R}\text{ }\rightarrow \Bbb{R)}.  \label{28}
\end{equation}

(The following considerations apply equally to the general relativistic H-J
equation,

\[
g^{ab}\partial _{a}S\partial _{b}S+m^{2}=0.) 
\]

We use the existence of an $n$ parameter family, ($s^{a}),$ of solutions to
the H-J equation, a ``complete solution'', \qquad \qquad \qquad 
\[
S=S^{*}(q^{a},s^{b},t), 
\]
i.e., a solution depending on $n$ parameters $s^{b},$ such that the equation

\[
\frac{\partial S^{*}}{\partial s^{a}}=\alpha _{a} 
\]
can be solved uniquely with respect to the variables $q^{a}.$

We now define what is to be our generating family, the function $\frak{F(}%
q^{a},s^{b},t)$ by$\qquad \qquad \qquad $%
\begin{equation}
\frak{F}=S^{*}(q^{a},s^{b},t)-F(s^{b})  \label{29}
\end{equation}
where $F$($s^{b}$) is an arbitrary function such that $\frak{F}$ obeys the
rank condition. We now require, from the theory of generating families, the
extremal condition, i.e., Eq.(\ref{C}), that \qquad \qquad \qquad \qquad 
\begin{equation}
\partial \frak{F/}\partial s^{b}\equiv \partial S^{*}\frak{/}\partial s^{b}-%
\partial F\frak{/}\partial s^{b}=0.  \label{30}
\end{equation}
From the general theory we infer that Eq.(\ref{30}) can be solved for either 
$s^{b}$ or $q^{a}$ or some combination of them or alternatively it allows us
to describe the solution parametrically, i.e.,

\[
s^{a}=S^{a}(y^{b},t),\text{ }q^{a}=Q^{a}(y^{b},t).\text{ } 
\]

Moreover, the resulting equations

\begin{eqnarray*}
q^{a} &=&Q^{a}(y^{b},t), \\
p_{a} &=&\frac{\partial }{\partial q^{a}}\frak{F}
(q^{a},s^{b},t)=P_{a}(y^{b},t),
\end{eqnarray*}
define a one-parameter family of Lagrangian submanifolds of $T^{*}M$,
parametrized by $t.$

They also define a Lagrangian submanifold of the phase space over $M$x$\Bbb{%
\ R=}\{q^{a},t\}$ with canonical coordinates ($q^{a},t,p_{a},-E),$ contained
in the ``physical hypersurface'' (constraint submanifold, $\frak{E}$ ) given
by $E=H(q^{a},p_{a}).$ A ``classical'' solution $S(q^{a},t)$ to the H-J
equation may be geometrically characterized as the generating function of a
Lagrangian section of $T^{*}(M$x$\Bbb{R)}$ contained in $\frak{E}$.
Therefore, it is reasonable to call any generating function of any
Lagrangian submanifold $L,$ contained in $\frak{E,}$ $[$explained earlier in
connection with generating families, see Eq.(\ref{II})] a ``generalized
solution'' of the H-J equation since it extends thru singularities the
Lagrangian submanifold defined (locally) by a ``classical'' solution $S$.
[Such an extension is unique since $L$ is ruled by phase trajectories
determined by Hamilton's equations and initial conditions.]

Note that indeed a generating function of this type does satisfy the
generalized H-J equation;

\[
H(q^{a},p_{a})\equiv H(q^{A},q^{J},p_{A},p_{J})\Rightarrow H(q^{A},-\partial
^{J}K,\partial _{A}K,p_{J})+\frac{\partial K}{\partial t}=0, 
\]
where we have used Eq.(\ref{22}).

In this sense, the construction, via Eqs.(\ref{29}) and (\ref{30}), provides
``generalized'' solutions of the H-J equation in that it extends an ordinary
solution thru singular points\cite{HKP}. It is also a ``general'' solution,
in the sense that by a suitable choice of a complete solution $%
S^{*}(q^{a},s^{b},t))$ of the H-J equation and the function $F(s^{b}),$ it
can be adapted to any Cauchy data. If one begins with arbitrary Cauchy data, 
$S_{0}(q^{a}),$ it is possible to convert it into an expression for $%
F(s^{b}).$ See the corresponding argument for the eikonal equation in Sec.IV.

This procedure allows us to choose an ensemble (of non-interacting particles
in three space) for a classical system and then see how the entire ensemble
evolves and study the density waves; for example the high density in the
neighborhood of caustics. See Sec.III. $\spadesuit $

$\spadesuit $An example closely related to the preceding one is that of the
eikonal equation in an arbitrary four dimensional Lorentzian manifold,
namely\qquad 
\begin{equation}
g^{ab}(x^{c})\partial _{a}S\partial _{b}S=0,
\end{equation}
where $g^{ab\text{ }}$is the inverse metric$\ $and $x^{c}$ are some local
coordinates. This time we start from a two parameter family of
solutions,\qquad \qquad \qquad 
\begin{equation}
S=Z(x^{c},\zeta ,\overline{\zeta })
\end{equation}
with ($\zeta ,\overline{\zeta })$ the parameters, chosen as the complex
stereographic coordinates on $S^{2}$ in order to label the sphere of null
directions at the space-time point $x^{c}.$ (We write expressions in terms
of both ($\zeta ,\overline{\zeta })$ in order to point out that the
functions used are not holomorphic in $\zeta $. Also it is convenient to
employ the independent directional derivatives with respect to ($\zeta ,%
\overline{\zeta }).)$ It is often difficult or even practically impossible
to find such solutions though there are perturbation techniques to construct
approximations to such solutions.

We now define our generating family by\cite{SGN1,SNG2}

\begin{equation}
\frak{F(}x^{c},\zeta ,\overline{\zeta })=Z(x^{c},\zeta ,\overline{\zeta }
)-F(\zeta ,\overline{\zeta })  \label{31}
\end{equation}
with $F(\zeta ,\overline{\zeta })$ an arbitrary function. (Often $F$ is
chosen as a regular function on $S^{2},$ though this is not necessary.) The
extremal condition, Eq.(\ref{C}), is now 
\begin{equation}
\partial \frak{F/\partial }\zeta =0,\text{ }\partial \frak{F/\partial }%
\overline{\zeta }=0.  \label{32}
\end{equation}
If these equations can be solved by $\zeta =\zeta (x^{a}),$ then $\frak{F[}%
x^{a}]=\frak{F(}x^{c},\zeta (x^{a}),\overline{\zeta }(x^{a}))$ also
satisfies the eikonal equation. (See Sec. IV for the flat -space version of
this with three parameters.) Note that this construction \textit{does not }
allow the construction of the general solution; to do that $F(\zeta ,%
\overline{\zeta })$ would have to depend on three parameters rather than
two. However this procedure does allow the construction of any arbitrary
single null hypersurface\cite{SNG2}. The first example of this section is an
illustration of this construction. A much more valuable example is the
construction\cite{SNlens} of the light-cone of some arbitrary but fixed
space-time point, x$_{0}^{a}$. This can be used to generalize the usual
treatment of gravitational lensing. In fact in this case, Eq.(\ref{33})
below, becomes the lens equation when two of the coordinates, the radial
distance from an observer and the time, are held constant\cite{Stachel}.

If $\frak{F}$ is held constant (say zero) then the three equations, Eqs.(\ref
{31}) and (\ref{32}), can be solved for \textit{three,} ($x^{i}),$ of the
four coordinates, x$^{a},$ in terms of the fourth one, $x^{*},$ i.e., they
have the form 
\begin{equation}
x^{i}=X^{i}(x^{*},\zeta ,\overline{\zeta }).  \label{33}
\end{equation}
Eq.(\ref{33}) represents the set of null geodesics that generate the big
wavefront, $\frak{F}=0$; they are labeled by the ($\zeta ,\overline{\zeta })$
and parametrized by one of its coordinates, $x^{*}$.$\spadesuit $

$\spadesuit As$ a final example, we show how the N-plane lens map (in the
standard weak field, thin lens approximation) of gravitational lens theory
can be obtained naturally via a generating family, as a Lagrangian
projection.

Suppose that the lightrays emitted from some point source $\frak{s}$ on a
source plane P are consecutively gravitationally deflected by N thin mass
distributions, M$_{i}$ before they reach an observer $\frak{O.}$ The M$_{i}$
are represented by surface mass densities in N planes P$_{i}$ orthogonal to
a straight line going thru $\frak{O}$ and perpendicular to the source plane
P. A virtual light path is represented as a polygon figure from $\frak{O}$
to $\frak{s}$ with vertices on the P$_{i}.$ The influence of the M$_{i}$ on
light can be expressed in terms of two-dimensional potentials $\Psi _{i}$ on
the planes P$_{i}.$ The travel time of a lightray depends not only on its
geometrical path length, but also on the gravitational Shapiro-time delay
suffered when the rays passes a ``lens'' M$_{i}.$ If the positions \textbf{s}%
$_{i}$ of a virtual ray on the plane P$_{i}$ and the position \textbf{q}$%
\equiv $\textbf{s}$_{N+1}$ of $\frak{s}$ on a source plane P$,$ are scaled
suitably, the (variable part) of the travel time has the form\cite{42}

\begin{equation}
\frak{F(}\text{\textbf{s}}_{1}\text{,\textbf{s}}_{2}\text{....,\textbf{s}}%
_{N},\text{\textbf{q}}\mathbf{)=\ }\sum_{i=1}^{N}C_{i}[\frac{1}{2}(\text{%
\textbf{s}}_{i}-\text{\textbf{s}}_{i+1})^{2}-\beta _{i}\Psi _{i}(\text{%
\textbf{s}}_{i})]  \label{VIa}
\end{equation}
where the constants $C_{i},\beta _{i}$ depend on the distances of $\frak{s}$
and the M$_{i}$ from $\frak{O.}$ Fermat's principle (which singles out the
``real'' from the virtual rays) in this idealization takes the form

\begin{equation}
\frac{\partial \frak{F}}{\partial \text{\textbf{s}}_{i}}=0.  \label{VIb}
\end{equation}

We consider $\frak{F(}$\textbf{s}$_{1},$\textbf{s}$_{2}....$\textbf{s}$_{N},$%
\textbf{q}$\mathbf{)}$ as a generating family with - in the notation of
Sec.V -

\[
s^{J}=(\text{\textbf{s}}_{1},\text{\textbf{s}}_{2}....,\text{\textbf{s}}
_{N}),\qquad q^{a}=\text{\textbf{q}}=\text{\textbf{s}}_{N+1}, 
\]
i.e., in this case we have $n=2$ and $m=2$N. The rank condition is
satisfied; indeed the solution from Eq.(\ref{VIb}) has the form

\[
\text{\textbf{s}}_{2}=f_{2}(\text{\textbf{s}}_{1}),\quad \text{\textbf{s}}%
_{3}=f_{3}(\text{\textbf{s}}_{2},\text{\textbf{s}}_{1})=f_{3}[\text{\textbf{s%
}}_{1}],\quad ..............,\quad \text{\textbf{q}}=f_{N+1}(\text{\textbf{s}%
}_{1}). 
\]
If we put 
\[
\mathbf{p}=\frac{\partial \frak{F}}{\partial \text{\textbf{q}}}=P(\mathbf{s}%
_{1}) 
\]
then according to the general theory the equations

\[
\text{\textbf{q}}=f_{N+1}(\text{\textbf{s}}_{1}),\quad \text{\textbf{p}}=P(%
\text{\textbf{s}}_{1}) 
\]
describe a Lagrangian submanifold of $T^{*}\Bbb{R}^{2}=\{\mathbf{q,p}\},$
parametrized by the ray direction (corresponding to \textbf{s}$_{1}$) at the
observer $\frak{O.}$ The associated Lagrangian projection is given by the
lens map

\[
\mathbf{s}_{1}\mapsto \text{\textbf{q}}=f_{N+1}(\text{\textbf{s}}_{1}) 
\]
which takes a ray direction \textbf{s}$_{1}$ at $\frak{O}$ to the source
position \textbf{q}$\mathbf{.}$ (Note that this is a gradient map if N$=1$
but a more general Lagrangian map for N $>1.)$ Critical curves, caustics,
types of singularities then can be analyzed according to the general theory.$%
\spadesuit $

We mention, with no discussion, that Lagrangian submanifolds play a role in
the characterization and construction of physical states of (linear) quantum
fields on (classical) curved, globally hyperbolic space-times $(M,g)$. Such
(Hadamard) states can be characterized by the ``wave front sets'' of their
two-point ``functions'', subsets of $T^{*}(M$x$M{)}${\cite
{Radzikowski,Junker}} which have been shown to be contained in Lagrangian
submanifolds of $T^{*}(M$x$M)$.

\section{Epilogue}

The study of caustics and wave front evolution has a rich history; it dates
back to the early studies of Newton and Huygens, Cayley studied the normal
wavefront evolution from the triaxial ellipsoid in the middle of the 19th
century. The contemporary study of generic wavefront and caustic behavior
arose in the mid-century via the classification studies of the singularities
of functions and mappings. It arose mainly via the efforts of the
mathematicians, H. Whitney, R. Thom and V.I. Arnold; the work of the latter
on Lagrangian maps has been the main concern here. With several notable
exceptions, in particular M. Berry and Ya. Zeldovich, physicists seem to
have largely ignored the subject even though it has implications for a wide
range of physical applications; all forms of wave propagation, both
classical and quantum mechanical; from geometric optics thru to physical
optics; intensity distributions in interference and diffraction phenomena 
\cite{MBII} (e.g., evaluations of the Fresnel and Airy integrals);
gravitational lensing; structure formation in the early universe and in
galaxies via density waves; finite size image disruption\cite{MBIII};
Hamilton-Jacobi theory; stability problems; thermodynamics; elasticity
theory; and states of quantum fields in curved space-times\cite{Junker}.

We have only attempted to give the most rudimentary treatment of the basic
mathematical ideas that lie at the origin of this large subject and to
introduce several potential applications of the general theory to physics.
For a variety of reasons we have avoided completely several relevant topics,
e.g., global topological questions, the theory of classification of critical
points of functions, the surprising relationship between the classification
of functions and the Weyl groups, etc.

There are several articles and books which contain extensive bibliographies
and historical surveys of the origin and development of singularity theory.
The book \textit{Catastrophe Theory\cite{31},} besides being a wonderful
introduction to the subject, contains both a brief history and an extensive
annotated bibliography to both the theory and its many applications. The
article in Russian Math. Surveys\cite{RM1}, dedicated to Arnold on his 60th
birthday, contains a complete list of Arnold's publications while Arnold's
article\cite{Arnoldp1}, on large scale issues in wave propagation (and as a
delightful aside, a discussion on Mathematics Education), also has a large
bibliography as does Arnold's article\cite{A6} in Vol.V in Dynamical Systems
of the Enc. of Math. Sciences. Though Arnold's book\cite{HB} ``\textit{\
Huygens \& Barrow, Newton \& Hooke}'' only touches on the details of
singularity theory it must be mentioned for its wealth of fascinating
historical observations. We point out that though most of our references are
to books and articles published later than 1980 almost all of the
fundamental mathematical work was completed by the mid 1970's. We list
several of the principle early references\cite{A7,A8,A9,A10}.

\qquad *************************************************

\ \qquad \qquad \textbf{V}.\textbf{I. Arnold on Mathematicians}

\noindent ``It is almost impossible for me to read contemporary
mathematicians who instead of saying

\qquad `Petya washed his hands' write simply

\noindent `There is a $t_{1}<0$ such that the image of $t_{1}$ under the
natural mapping $t_{1}\mapsto $ Petya($t_{1})$ belongs to the set of dirty
hands and a $t_{2},$ $t_{1}<t_{2}<0,$ such that the image of $t_{2}$ under
the above mentioned mapping belongs to the complement of the set defined in
the preceding sentence' ''.

\section{Acknowledgments}

ETN thanks Simonetta Frittelli and Carlos Kozameh for the many stimulating
and enlightening ``fights'' and discussions which played a major role in
clarifying his understanding of this subject. He also thanks the NSF for
support under research grants \# PHY 92-05109 and PHY 97-22049. JE is
grateful to Peter Schneider and his group for introducing him to lensing, to
Helmut Friedrich for clarifying remarks on the theory of wavefronts and to
Thomas Buchert for sharing with him thoughts on applications of singularity
theory to cosmology.

We extent a special thanks to V.I. Arnold for his help in clarifying certain
historical issues and for help in greatly expanding our bibliography.

In particular we offer an extra special thanks to Simonetta Frittelli for
the preparation of the figures used here.

\end{document}